\documentclass[a4paper,11pt]{article}

\usepackage[dvips]{graphicx}
\usepackage{amsfonts, color, float,  amsmath}
\usepackage{amssymb}\usepackage{subfigure}
\usepackage{url}
\usepackage{array}

\usepackage{hyperref}

\usepackage{slashed}
\usepackage[dvipsnames]{xcolor} 

\usepackage[small,bf,sf]{caption} 
\usepackage{microtype}
\usepackage{cite}
\usepackage{authblk}
\usepackage{datetime}
\usepackage[bottom]{footmisc} 
\usepackage{booktabs}  

\usepackage{colortbl}
\definecolor{myblue}{rgb}{0.8,0.85,1} 
\definecolor{light-gray}{gray}{0.95}

\newdateformat{mydate}{\THEDAY \  \monthname \  \THEYEAR}

\allowdisplaybreaks

\newcommand{\MSb}{{\overline{\rm MS}}}
\newcommand{\wh}{\widehat}

\newcommand{\ket}[1]{\left|#1\right\rangle}

\def\beq {\begin{equation}}
\def\eeq {\end{equation}}
\def\bea {\begin{eqnarray}}
\def\eea {\end{eqnarray}}

\def\nn {\nonumber}

\def\epem{e^+e^-\to(\mbox{hadrons})}

\def\Dlog{{\rm Dlog}}
\def\PDlog{\mathbb{D}{\rm log}}

\def \wh{\widehat}

\AtBeginDocument{}

\title{\LARGE {\bf \sffamily \boldmath  Higher-order QCD corrections to hadronic $\tau$ decays  from Pad\'e approximants}}

\author[a]{Diogo Boito}
\author[b]{Pere Masjuan}
\author[a]{Fabio Oliani\vspace{0.3cm}}

\affil[a]{\it  Instituto de F\'isica de S\~ao Carlos, Universidade de S\~ao Paulo, CP 369, 13560-970, S\~ao
Carlos, SP, Brazil\vspace{0.3cm}}

\affil[b]{\it Grup de F\'{\i}sica Te\`orica, Departament de F\'{\i}sica, 
  Universitat Aut\`onoma de Barcelona,  and Institut de F\'{\i}sica d'Altes Energies (IFAE), The Barcelona Institute of Science and Technology (BIST), 
  Campus UAB, E-08193 Bellaterra (Barcelona), Spain }

\date{}

\begin{document}

\begin{flushright}
{\mydate \today}
\end{flushright}

\vspace*{-0.7cm}
\begingroup
\let\newpage\relax
\maketitle
\endgroup
\date{}

\begin{abstract}
  \noindent
  Perturbative QCD corrections to hadronic $\tau$ decays and $e^+e^-$ annihilation
into hadrons below charm are obtained from the  Adler
function, which at present is known  in the chiral limit to five-loop accuracy. Extractions
of the strong coupling, $\alpha_s$, from these processes suffer from an ambiguity related to
the
treatment of unknown higher orders in the perturbative series. In this
work, we exploit the method of Pad\'e approximants and its convergence theorems to extract
information about higher-order corrections to the Adler function in a systematic way. First, the method is
tested in the large-$\beta_0$ limit of QCD, where the perturbative
series is known to all orders. We devise strategies to accelerate the
convergence of the method employing renormalization scheme variations
and the so-called D-log Pad\'e approximants. The success of these
strategies can be understood in terms of the analytic structure of the series in the Borel plane.
We then apply the method
to full QCD and obtain  reliable model-independent predictions for the higher-order
coefficients of the Adler function. For the six-,
seven-, and eight-loop coefficients we find $c_{5,1} = 277\pm 51$, $c_{6,1}=3460\pm 690$, and $c_{7,1}=(2.02\pm0.72)\times 10^4$, respectively,
with errors to be understood as lower and upper bounds.
Our model-independent reconstruction of the
perturbative QCD corrections to the $\tau$ hadronic width strongly
favours the use of fixed-order perturbation theory (FOPT) for the
renormalization-scale setting.

\end{abstract}

\thispagestyle{empty}


\newpage
\tableofcontents

\setcounter{page}{1}


\section{Introduction}

The precise determination of the strong coupling, $\alpha_s$, is a key
ingredient for calculations of all processes involving perturbative
Quantum Chromodynamics (QCD) and represents a fundamental test of the
internal consistency of the Standard Model. The value of $\alpha_s$, together with the mass of the top quark, plays, for example, a
crucial role in the fate of the Standard Model vacuum~\cite{SMVacuum}.
The extraction of $\alpha_s$ from hadronic $\tau$
decays~\cite{BNP92,Boito3, PRS16,ALEPH} (and also from $e^+e^-\to
(\mbox {hadrons})$ below charm~\cite{KNT18,epem})  is of special interest for two
reasons. First, because it is  done at relatively low energies, close
to the limit of validity of perturbative QCD. Therefore, the evolution
of $\alpha_s$ from the $\tau$ mass scale to the $Z$ mass scale represents
one of the most non-trivial tests of asymptotic freedom~\cite{PichTP}
as predicted by the QCD
$\beta$-function~\cite{QCDBeta4,QCDBeta5,Herzog:2017ohr,Luthe:2017ttg}. Second, this determination
of $\alpha_s(m_Z)$ is competitive, since the running reduces the size
of the relative error.

However, theoretical uncertainties still affect
the determination of $\alpha_s$ from these processes.
At and around the $\tau$ mass, perturbative QCD is
still valid, but non-perturbative effects become non-negligible.
These effects are smaller by a factor of about ten when compared to
the perturbative QCD contribution, but must be taken into account
carefully~\cite{Boito3,CriticalAp}. They are encoded in the condensates of the Operator
Product Expansion (OPE) and the related contributions from violations of
quark-hadron duality --- or simply {\it duality violations}
(DVs)~\cite{Shifman,CGP2005,DVsMainz,DVs2018}.

Another important source of uncertainty stems from the
renormalization-scale setting in the perturbative contribution.
Theoretically, the decay $\tau \to ({\rm hadrons}) +\nu_\tau$ is
expressed in terms of a weighted integral of the hadronic spectral
functions that runs over the hadronic invariant mass squared from
threshold up to $m_\tau^2$.  Since perturbative QCD cannot be trusted
at low energies, one resorts to Finite Energy Sum Rules (FESRs) to
relate this integral to an integral along a closed contour in the
complex plane with $|s| = m_\tau^2$. In this process, a procedure must
be adopted to set the renormalization scale. The two most commonly
employed procedures are known as Fixed Order Perturbation Theory
(FOPT) \cite{BJ08} and Contour Improved Perturbation Theory (CIPT)
\cite{PLD92,Pivovarov:1991rh} (they are discussed in more detail below). The two
represent different ways of treating the unknown higher orders in
perturbation theory, and lead to different perturbative series and,
therefore, to different values of~$\alpha_s$. This remains true in
the analogous extraction of $\alpha_s$ from $e^+e^- \to (\mbox{hadrons})$ below charm although,
numerically, the difference between the two procedures is smaller in
that case~\cite{epem}.

The elimination of this ambiguity is inherently difficult because it
requires knowledge about higher orders of the perturbative expansion.
At present, the perturbative QCD expansion of the Adler function in the chiral limit is known up to $\alpha_s^4$ thanks to
the five-loop computation of Ref.~\cite{BCK08,Herzog:2017dtz} and it is unlikely that
the result at six loops will be available anytime
soon~\cite{JKComm}. In the absence of exact calculations for the
higher-order coefficients, one must tackle this problem with methods
that allow for a partial reconstruction of the series based only on
the available information.

The general structure of the perturbative series is assumed to be
known. It is an asymptotic series (therefore divergent) with
coefficients that grow factorially. The divergent behaviour of the
series is governed by {\it renormalons}: singularities along the real axis of the  Borel transformed series~\cite{Renormalons}. The
position of these singularities is known since they are related to
the dimension of operators that participate in the OPE of the relevant correlator. The exponents of the
singularities are related to the anomalous dimension of the same operators
and can, in principle, be calculated. On the other hand, nothing,
essentially, is known about the residues of the singularities.

This partial knowledge has been used to construct realistic
representations of the full series by approximating its Borel
transform, which has an infinite tower of singularities, by a small
number of dominant ones~\cite{BJ08,BBJ13}.  These models for the Borel
transformed series are, in some  cases, a type of {\it rational} or
{\it Pad\'e} approximant.\footnote{To be precise, the models of
  Refs.~\cite{BJ08,BBJ13} are akin to Pad\'e-type
  approximants~\cite{Pades,MP07}.} Motivated by this observation, in
this paper we investigate, systematically, the use of Pad\'e
theory~\cite{Pades,MP07,MP08,MP09} to reconstruct the Adler function,
which  governs hadronic $\tau$ decays and the cross
section of $\epem$.  One of the main advantages of the use of rational
approximants, as compared to the so-called renormalon models of
Refs.~\cite{BJ08,BBJ13}, is that they can be made model
independent. Additionally, in some well defined cases, theorems guarantee the
convergence of a sequence of approximants to the function of
interest. 

In the past, rational approximants have already been used in the
context of $\tau$ decays~\cite{SEK95}. In particular, the observation
that their convergence can be improved when one uses the Borel transformed
series, as opposed to the series in $\alpha_s$, was already made.
(This procedure is sometimes referred to as ``Pad\'e-Borel
method''~\cite{BorelPade}.) At that time, however, the
perturbative series was known to one order less, only up to
$\alpha_s^3$. Moreover, the connection with renormalons in
applications to $\tau$ decays was not made explicitly\footnote{A
  discussion of renormalons and Pad\'e approximants does appear in the
  context of the Bjorken sum-rule in a related
  paper~\cite{EGKS96}.}. Here, this connection is established and we
are able to use different types of Pad\'e approximants (such as
partial Pad\'e approximants) thanks to the available knowledge about the
renormalon singularities.

To validate our approach, before applying Pad\'e approximants (PAs) to
full QCD, we will work within the large-$\beta_0$ limit. In this
limit, one obtains all the corrections with highest power of $N_f$ at
every given $\alpha_s$ order in the perturbative expansion (with $N_f$
being the number of light-quark flavours).  The application of this
procedure to $\tau$ decays generates an asymptotic series in
$\alpha_s$ that is known to all orders, and the result for the Borel
transformed Adler functionn can be written in a compact
form~\cite{MB93,DB93}. This Borel transform is a meromorphic function
in the complex plane: it has a finite radius of convergence and an
infinite number of renormalon poles along the real axis, but no branch cut.
Therefore, the theory of PAs to meromorphic functions, and in
particular Pommerenke's theorem, apply~\cite{Pades,Pommerenke}. Apart
from the standard PAs, we will consider several strategies for
accelerating the convergence of the approximation. First, we will
exploit the renormalization scheme dependence of the perturbative
series, following Ref.~\cite{BJM16}, to optimize the convergence of
the Pad\'e approximants to exactly known Adler function. Additionally,
we will consider partial Pad\'e approximants (which exploit the
available knowledge about the renormalon singularities). We will also
employ D-log Pad\'e approximants~\cite{Pades} which can also be very
effective in approximating functions with branch cuts.  Finally, we
investigate the application of the different PAs to the FOPT expansion
of the QCD corrections to the $\tau$ hadronic width. This series has a
much simpler analytic structure in the Borel plane, which leads
to coefficients that follow a more regular pattern, and is more amenable
to approximation by rational functions. From these approximants one
can easily perform an indirect reconstruction of the Adler function that
is very reliable and requires little information.

 The systematic study performed in large-$\beta_0$ and the lessons we
 learn from this limit are used as the basis for the QCD analysis.  In
 the case of full QCD, the structure of the Borel transformed Adler
 function is more involved, since the poles become branch
 cuts~\cite{Renormalons}. From renormalization scheme variations we
 find indications that in QCD the leading UV renormalon is suppressed
 with respect to large-$\beta_0$ and, as a consequence, the sign
 alternation of the series is probably postponed. We will then show
 that, as in large-$\beta_0$, it is advantageous to consider Pad\'e
 approximants to the FOPT expansion of the corrections to the $\tau$
 hadronic width. From these approximants, we are able to obtain
 reliable model-independent predictions for the higher-order
 coefficients of the Adler function, together with an estimate of
 their uncertainty, and extract an estimate for the ressumed value of
 the perturbative QCD corrections to hadronic $\tau$ decays. We are
 also in a position to discuss the renormalization group improvement
 of the series and we show, from our reconstruction of high orders,
 that FOPT is strongly favoured in QCD. The systematic use of PAs lead
 to results that are model independent and that have significantly
 smaller errors than results obtained from other methods.

Our work is organized as follows. In Sec.~\ref{sec:taudecays} we
describe the essentials about the QCD description of hadronic $\tau$
decays and, in Sec.~\ref{sec:Pades}, we collect the main facts about
Pad\'e theory relevant to this work. Then, in
Sec.~\ref{sec:largebeta}, we apply the Pad\'e approximants to the
large-$\beta_0$ limit of QCD.  The results in full QCD are presented
in Sec.~\ref{sec:QCD} and the conclusions are given in
Sec.~\ref{sec:conclusions}.


\section{\boldmath Perturbative QCD in hadronic $\tau$ decays}
\label{sec:taudecays}

In the study of perturbative QCD corrections to hadronic $\tau$ decays
and $\epem$ below charm the central object is the Adler function in
the chiral limit (defined below).  From the knowledge of its expansion
 one can derive the corrections to the $\tau$
hadronic width~\cite{BJ08} as well as the perturbative expansion of
$R(s)$ for $\epem$ below charm~\cite{epem}.  In the spirit of being
self-contained, we review here the main aspects of the theoretical
description of hadronic $\tau$ decays in QCD. (We refer to
Refs.~\cite{BNP92,BJ08, alphas2011} for further details.) Although we
frame our discussion in the context of $\tau$ decays the application
to $\epem$ is straightforward and is, essentially, a matter of
normalization to reflect the fact that the current in that case is the
electromagnetic one. For the details regarding this normalization we
refer to Ref.~\cite{epem}.

The decay rate of $\tau \to ({\rm hadrons})+\nu_\tau$ can be separated
experimentally into three components: a vector and an axial-vector, due
to decays mediated by the light-quark $\bar u d$ current, and strange
contributions, arising from the $\bar u s$ current. These decay rates,
normalized to the width of $\tau \to e \bar \nu_e \nu_\tau$,
are denoted $R_{\tau,V}$, $R_{\tau,A}$, and $R_{\tau,S}$, respectively. When
extracting $\alpha_s$ it is convenient to work only with light quarks,
because corrections proportional to the mass of the quarks can then be
safely neglected. Here, we restrict ourselves to the non-strange
channels, precisely for this reason. Then, the  
different corrections to the partonic result can be parametrized as
\beq
R_{\tau,V/A} = \frac{N_c}{2}S_{\rm EW} |V_{ud}|^2 \left[1 + \delta^{(0)} + \delta_{\rm NP}  +\delta_{\rm EW}\right],
\eeq
where $S_{\rm EW}$ and $\delta_{\rm EW}$ are  small  electroweak corrections and
$V_{ud}$ the CKM matrix element; the unity, in between square
brackets, is simply the partonic result. The first correction,
$\delta^{(0)}$, is the perturbative QCD part, which is the dominant
contribution ($\sim 20\%$). Non-perturbative contributions, encoded in $\delta_{\rm NP}$, are
significantly smaller and contain both OPE condensates and 
DVs. 
In this work, we focus on the perturbative QCD part, $\delta^{(0)}$,
that we discuss in more detail below.

The relevant quantity that governs $R_{\tau,V/A}$ are the correlators
\beq
\Pi_{V/A}^{\mu \nu}(p) \equiv i \int dx \,e^{i p x}\, \langle \Omega | T \{ J_{V/A}^{\mu}(x) J_{V/A}^{\nu}(0)^{\dagger} \} | \Omega \rangle,\label{eq:Corr}
\eeq
where $\ket{\Omega}$ represents the physical vacuum and the currents
are $J^\mu_{V/(A)}(x)=(\bar u \gamma^\mu (\gamma_5)d)(x)$.  These
correlators can be decomposed into transverse, $\Pi^{(1)}_{V/A}(s)$,
and longitudinal, $\Pi^{(0)}_{V/A}(s)$, parts in the usual way (with
$s=p^2$). The decay rate can be expressed in terms of integrals over the
spectral functions, $\frac{1}{\pi}{\rm Im}\Pi^{J}_{V/A}(s)$, that run
from $s=0$ to $s=m_\tau^2$~\cite{BNP92,BJ08}. These integrals, on the theory side, are problematic  because perturbative QCD is not valid at low energies. To
circumvent this problem, one resorts to a FESR that exploits the
analyticity properties of the correlators. The quantities
$R_{\tau,V/A}$ can then be expressed as an integral in a closed
contour in the complex plane with fixed $|s|=m_\tau^2$.
Explicitly, for the perturbative contribution, one finds~\cite{BJ08}
\beq
\delta^{(0)} = \frac{1}{2\pi i}  \oint\displaylimits_{|x|=1}\frac{dx}{x} W(x) \widehat D^{(1+0)}_{\rm pert}(m_\tau^2x),\label{eq:delta0}
\eeq
with  $x = s/m_\tau^2$ and where the weight
function $W(x)$, determined by phase space, is $W(x) = (1-x)^3(1+x)$.\footnote{Perturbative corrections to the spectral function are obtained simply by using $W(x)=1$ in Eq.~(\ref{eq:delta0}).}
In this integral, $ \widehat D^{(1+0)}_{\rm pert}$ is  the perturbative contribution to the reduced Adler function defined  as
\beq
1+ \widehat D_{\rm pert}(Q^2)
= \frac{12\pi^2}{N_c}D^{(1+0)}_{\rm pert}(Q^2),\label{RedAdler}
\eeq
where the Adler function itself, $D^{(1+0)}$, is obtained as the logarithmic derivative of
the combination $\Pi^{(1+0)}(s)$ as
\beq
D^{(1+0)}(s)= -s \frac{d}{ds}\left[ \Pi^{(1+0)}(s)\right].
\eeq
The Adler function is a physical object in the sense that it does not contain
subtraction constants that depend on the renormalization conventions.
This function is central to our work.

The perturbative expansion of $\widehat D$  starts at $\mathcal{O}(\alpha_s)$ and can be cast as 
\beq
\widehat D_{\rm pert}(s) =  \sum\limits_{n = 1}^{\infty}{a^{n}_{\mu}} \sum\limits_{k = 1}^{n+1 }{k c_{n,k}L^{k-1}},\label{eq:AdlerExp}
\eeq
where $L=\log(-s/\mu^2)$ and $a_\mu =\alpha_s(\mu)/\pi$.  In this
expansion, the only independent coefficients are the $c_{n,1}$; the
others can be obtained imposing renormalization group (RG) invariance,
and are expressed in terms of the $c_{n,1}$ and $\beta$-function
coefficients~\cite{BJ08,MJ05}. The logarithms  can be summed with the scale choice $\mu^2 = -s \equiv Q^2$ giving
\beq
\widehat D_{\rm pert}(Q^{2}) =  \sum\limits_{n = 1}^{\infty}{c_{n,1}a^{n}_{Q}} \equiv \sum_{n=0}^\infty r_n \alpha_s^{n+1}(Q).\label{eq:DCIPT}
\eeq
where $r_n=c_{n+1,1}/\pi^{n+1}$.
With this definition, the perturbative expansion of the reduced  Adler function with the choice $\mu^2=Q^2$  then  reads (for $N_f=3$, $\MSb$ scheme)\footnote{We will often drop the subscript ``pert'' in $\widehat D$.}
\beq
\widehat D(Q^2) = a_Q + 1.640 \, a_Q^2 +6.371\, a_Q^3+ 49.08\, a_Q^4+\cdots, \label{DinQCD}
\eeq
from which the numerical values of the known independent coefficients $c_{n,1}$ of Eq.~(\ref{eq:DCIPT})  can be immediately read off.


The perturbative series of Eq.~(\ref{eq:AdlerExp}) is divergent and
one assumes that it must be an asymptotic expansion to the true value
of the function being expanded~\cite{Renormalons}. To study the
perturbative contribution to the Adler function, and in particular its
renormalon content, it is therefore convenient to work with the Borel
transformed series, which can have a finite radius of convergence,
defined~as
\beq
B[\widehat D](t) \equiv \sum\limits_{n=0}^{\infty}{r_{n} \frac{t^{n}}{n !}}.
\label{BorelDef}
\eeq
The original expansion, in turn,  can be understood as an asymptotic series to the inverse Borel transform
\beq
\widehat D (\alpha) \equiv \int\limits_{0}^{\infty}{dt \text{e}^{-t/\alpha}B[\widehat D](t)},\label{BorelInt}
\eeq
provided that the integral exists. The last equation defines the Borel sum of the asymptotic series. The divergence of the original series, $\widehat D$, is translated into singularities in the $t$ variable. Two types can be
distinguished: ultraviolet (UV) and infrared (IR) renormalons.
 The UV renormalons lie on the negative real axis and contribute
with sign alternating coefficients. IR renormalons are singularities on the positive real axis which contribute with
fixed sign coefficients. The latter  obstruct the integration in Eq.~(\ref{BorelInt}) and generate an
ambiguity in the inverse Borel transform which is expected to cancel against power corrections of the OPE. The position of the singularities in the $t$ plane can be determined with general renormalization group (RG) arguments. They appear at
positive and negative integer values of the variable $u\equiv \frac{\beta_1 t}{2\pi}$ (except for $u=1$), where $\beta_1$ is the leading coefficient of the QCD $\beta$-function.\footnote{We define the QCD $\beta$-function as in Ref.~\cite{BJ08}
\[
\beta(a_\mu) \equiv -\mu \frac{d a_\mu}{d\mu} = \beta_1a_\mu^2 +\beta_2a_\mu^3+\beta_3a_\mu^4 + \beta_4a_\mu^5+\beta_5a_\mu^6+\cdots
\]}  The UV renormalon at
$u=-1$, being the closest to the origin, dominates the large order behaviour of the series, which must, therefore, be sign alternating at  higher orders. As seen in Eq.~(\ref{DinQCD}), this sign alternation is still not apparent   in the first four coefficients of the QCD expansion in the $\MSb$ scheme, which are known exactly.

Finally, to obtain the perturbative corrections to $R_{\tau,{V/A}}$ one needs to perform the integral in Eq.~(\ref{eq:delta0}). In the process,
one  needs to adopt a procedure in order to set the scale $\mu$, which
enters, implicitly, through Eq.~(\ref{eq:AdlerExp}). A
running scale, $\mu^2=Q^2$, as done in Eq.~(\ref{eq:DCIPT}), gives rise to
the aforementioned Contour-Improved Perturbation Theory (CIPT), where the running of
$\alpha_s$ along the contour is resummed to all orders. In this case, $\delta^{(0)}$ can
be written as
\beq
\delta^{(0)}_{\rm CI} = \sum_{n=1}^{\infty} c_{n,1}J_n^{\rm CI}(m_\tau^2), \qquad \mbox{with}  \qquad J_n^{\rm CI}=  \frac{1}{2\pi i}\oint\displaylimits_{|x|=1} \frac{dx}{x}(1-x)^{3}(1+x)a^{n}(-m_\tau^2x).
\eeq
Another option is to employ a fixed scale
$\mu^2=m_\tau^2$, which gives rise to  Fixed Order Perturbation
Theory\footnote{Here we will consider only CIPT and FOPT, but alternative schemes for setting the scale $\mu$ have been advocated in the literature~\cite{Caprinietal1,Caprinietal2,Caprinietal3,Caprinietal4,CLMV10}.}. Then, because $\alpha_s$ is evaluated at a fixed
scale, it can be taken outside the contour integrals, which are performed over
 the logarithms that appear in Eq.~(\ref{eq:AdlerExp}) as
\beq
\delta^{(0)}_{\rm FO} = 	\sum\limits_{n=1}^{\infty}{a_{\mu}^{n}} \sum\limits_{k=1}^{n}{k c_{n,k}} J^{\rm FO}_{k-1},\qquad \mbox{with} \qquad J_{n}^{\rm FO} \equiv \frac{1}{2 \pi i}\oint\limits_{|x|=1}{\frac{dx}{x}(1-x)^{3}(1+x)\ln^{n}(-x)}.\label{FOPTdef}
\eeq
Therefore, $\delta^{(0)}_{\rm FO}$ can also be written as an expansion in the coupling
\beq
\delta^{(0)}_{\rm FO} = \sum_{n=1}^\infty d_n a^n_Q,\label{FOPTexp}
\eeq
where the coefficients $d_n$ depend then on $c_{n,1}$, on the $\beta$-function coefficients, and on the integrals $J^{\rm FO}_n$. In QCD, this expansion reads, for $N_f=3$ and in the $\MSb$ scheme,
\beq
\delta^{(0)}_{\rm FO} = a_Q + 5.202\,  a_Q^2 + 26.37\, a_Q^3 + 127.1 \,a_Q^4 +( 307.8+c_{5,1})\,a_Q^5 + \cdots \label{FOPTQCD}
\eeq
where we give the numerical result of the known contributions to the first unknown coefficient.

In $\tau$ decays, using $\delta^{(0)}_{\rm CI}$ to extract
$\alpha_s(m_\tau^2)$ one obtains results about $5\%$ larger than those
obtained from $\delta^{(0)}_{\rm FO}$~\cite{Boito3}. (This difference
is reduced to about $2\%$ when $\alpha_s(m_\tau^2)$ is extracted from
$e^+e^-\to (\mbox{hadrons})$~\cite{epem}.) The elimination of this
ambiguity would therefore contribute to the extraction of $\alpha_s$
around $m_\tau^2$ with smaller uncertainties.


\section{Elements of Pad\'e theory}
\label{sec:Pades}

Let us consider a function $f(z)$ that assumes  a series expansion in the complex plane around $z=0$ 
\beq
f(z) = \sum_{n=0}^{\infty} f_n z^n.\label{fz}
\eeq
A Pad\'e approximant (PA) to $f(z)$~\cite{Pades}, denoted $P^M_N(z)$, is defined as the ratio of two polynomials in the variable $z$ of order $M$ and $N$, $Q_M(z)$ and $R_N(z)$, respectively, with the definition $R_N(0)=1$. This approximant is said to have a ``contact" of order $M+N$ with the expansion of the function $f(z)$ around the origin of the complex plane: the expansion of $P^M_N(z)$ around the origin  is the same as that of $f(z)$ for the first $M+N+1$ coefficients
\beq
P_N^M(z) = \frac{Q_M(z)}{R_N(z)} \approx f_0 +f_1\,z+f_2\,z^2+\cdots+ f_{M+N}z^{N+M} + \mathcal{O}\left( z^{M+N+1}  \right).
\eeq
From the reexpansion
of the approximant $P_N^M(z)$ one can read off an estimate for the
coefficient $f_{M+N+1}$, the first that is not used as input~\cite{MP08}. Estimates of this type will be of special
interest in this work.

The successful use of Pad\'e approximants to obtain quantitative
results about the function $f(z)$ requires only a qualitative
knowledge about the  analytic properties of the function. The PAs
can also be used to perform a reconstruction of the singularity
structure of $f(z)$ from its Taylor expansion. Convergence theorems
exist for the cases of analytic and single-valued functions with
multipoles or essential singularities~\cite{Pades}. Even for functions that
have branch points the PAs can be used, in many cases,
successfully. In these cases, for increasing order of approximation,
the poles of the PAs tend to accumulate along the branch cut,
effectively mimicking the analytic structure of the function~\cite{Pades}.

We will focus on sequences of Pad\'e approximants with $N= M+k$, for a
fixed value of $k$. For $k\neq 0$, the PAs $P_{M+k}^M$ define a {\it
  near diagonal} sequence while the case $k=0$ defines the diagonal
sequence. Pommerenke's Theorem~\cite{Pommerenke} then guarantees that
a sequence $P_{M+k}^M$ to a meromorphic function is convergent in any
compact set of the complex plane, except in a set of zero area that
includes the poles of the function $f(z)$, where even the original
function is not well defined. If there are other nuisance poles in the
approximant, the theorem requires that they move away from the region
as soon as $M$ grows, or appear in combination with a nearby zero,
which is called a {\it defect} or {\it Froissart
  doublet}~\cite{Pades}. In contrast, poles that are present in $f(z)$
tend to be relatively stable as one increases the order $M$.

In this paper, most of the times, the role of the function $f(z)$ is played  by the
Borel transform of the Adler function, defined in
Eq.~(\ref{BorelDef}). A key feature of the Borel transform, as already
discussed, is its singularities along the real axis, the
renormalons. It will be of interest to us to study how this
singularity structure is mimicked by the PAs. It is important to note that
when $f(z)$ is a general
meromorphic function some of the poles (and residues) of the
approximant $P_N^M(z)$ may become complex, even though the original
function has no complex poles.\footnote{When the meromorphic function is of the Stieltjes type the poles will always be  along the real axis. The functions we approximate in this work are not of this type. We will discuss this in more detail in the remainder.} Such poles cannot be identified with
any of the renormalon singularities, but they do not prevent the use
of $P_N^M(z)$ to study the function away from these poles. In fact, in the
process of approximating a function with an infinite number of poles
by an approximant that contains only a handful of them, the appearance of
these extraneous poles is expected to
happen~\cite{MP07}.

In the case at hand there is some available knowledge about the singularities of
the functions being approximated, which are precisely the renormalon
singularities of the Borel transformed Adler function. It may be desirable
to construct approximants that exploit this knowledge. If a set of poles
of the function are known, one can construct a so-called partial Pad\'e  approximants (PPA)~\cite{PPAs} defined as
\beq
\mathbb{P}_{N,K}^M(z) = \frac{Q_M(z)}{R_N(z) T_K(z)}.
\label{PartialPades}
\eeq
The polynomial $T_K(z)$, of order $K$, is constructed such as to have
$K$ zeros at the exact location of the first $K$ poles of $f(z)$. The
coefficients of the polynomials are again fixed by matching to the
Taylor expansion of $f(z)$, in the same way as for the PAs. Again, for
a general meromorphic function, complex-conjugated poles may appear in
the PPAs. The extreme case $N=0$ results in a Pad\'e-type approximant,
an approximant with the whole denominator given in advanced, less expensive
in terms of Taylor coefficients than a PA or a PPA.

The approximation of functions with branch points and cuts --- as is
the case for the Borel transform of the Adler function in QCD --- is
more subtle. In this case, a possible strategy is  the
manipulation of the series to a form which is more amenable to the  approximation by  Pad\'es. Let us consider the particular case of a function $f(z) =
\frac{A(z)}{(\mu-z)^{\gamma}} + B(z)$ with a cut from $\mu$ to
$\infty$ with exponent $\gamma$ and a reminder $B(z)$ with little
structure (both $A(z)$ and $B(z)$ are to be analytic at $z=\mu$). Following the method of Baker called D-log Pad\'e
approximant~\cite{Pades}, we can form PAs not to
$f(z)$ but to
\begin{equation}\label{Dlog}
F(z) = \frac{\rm d}{ {\rm d}z} \log[f(z)] \sim \frac{\gamma}{\mu-z} \quad \quad (\textrm{near } z=\mu)\, ,
\end{equation}
which turns out to be a meromorfic function to which the convergence theorem applies.
The use of appropriate Pad\'e approximants to $F(z)$ determines
in an unbiased way
both  the pole position, $z=\mu$, and the residue, $-\gamma$, which
  corresponds to the exponent of the cut of $f(z)$.
No assumption about neither $\mu$ nor $\gamma$ is made; they are determined
directly from the series coefficients. The approximation of $F(z)$ by a PA
yields an approximant for $f(z)$ that is not necessarily a rational function. To be more specific, the Dlog-PA approximant  to $f(z)$ obtained from using $P_N^M$ to approximate $F(z)$, that we denote $ {\rm Dlog}_N^M(z)$,  is 
\begin{equation}\label{DlogNM}
{\rm Dlog}_{ N}^M(z) = f(0) e^{\int d z \frac{Q_M(z)}{R_N(z)}}\, ,
\end{equation}
\noindent
where $P_N^M(z)= \frac{Q_M(z)}{R_N(z)} $ is the aforementioned PA to
$F(z)$. Due to the derivative in Eq.~(\ref{Dlog}), the constant
$f(0)$ is lost and must be reintroduced in order to properly normalize the
${\rm Dlog}^M_N(z)$. In practice, the non-rational approximant ${\rm Dlog}_{ N}^M(z)$ can  yield a rich analytical structure, in particular the presence of
branch cuts --- not necessarily present in the function $f(z)$ --- is to be expected.

 In case the branch point would
be known in advanced, one can form what we will call partial D-log Pad\'e approximants. They consist in forming  Pad\'e approximants to
\begin{equation}\label{DlogPP}
G(z) = (\mu-z) \frac{\rm d}{ {\rm d}z} \log[f(z)] \sim \gamma, 
\end{equation}
for an assumed value of $\mu$, which would yield a prediction for $\gamma$ by evaluation of the approximants around $z=\mu$. The approximant to $f(z)$ entailed by this procedure will be denoted $ \mathbb{D}{\rm log}^M_N(z;\mu)$ and is given by
\beq\label{DlogPPNM}
\mathbb{D}{\rm log}^M_N(z;\mu) = f(0) e^{\int d z \frac{Q_M(z)}{(\mu-z)R_N(z)}}.
\eeq
One
should remark that in Eqs.~(\ref{DlogPP}) 
no
assumption is made about $\gamma$. The method was originally designed
to be used in the presence of branch points, but if $\gamma$ is an
integer it can also work very well, as we show in Sec.~\ref{DlogSec}.

In summary, the Pad\'e approximants $P_N^M(z)$ can be viewed as an
economic and completely model-independent procedure, since all the
poles are left free and no analytical information about the
singular structure of the function needs to be included. They are,
however, expensive in terms of series coefficients. In order to fasten
the convergence range, the use of PPAs $\mathbb{P}_{N,K}^M$ improves
the results but requires knowledge about the singularities of the
function $f(z)$. Such singularities may be determine by PAs or by
external information.

D-log Pad\'e approximants, in turn, offer the possibility to exploit
the Pad\'e theory for functions with mutlipoles and branch cuts at the
expense of losing, due to the required derivative, the first
Taylor coefficient. Finally, the partial D-log Pad\'e approximant such
as Eq.~(\ref{DlogPPNM}), provides further improvement but
requires knowledge about the position of the singular
points. At the end of the day, for each case of interest, the PA
practitioner shall decide for the best strategy. As we will show in
the next section, a sequential study using the different
aforementioned approximations is the optimal way to extract
information about unknown Taylor coefficients and
about the singular structure of the objective function.


\section{\boldmath Pad\'e approximants in the large-$\beta_0$ limit}
\label{sec:largebeta}


A good laboratory for the strategy we present here is the so called
large-$\beta_0$ limit of QCD. Results in this limit are obtained
by first considering a large number of fermion flavours, $N_f$,
keeping $\alpha_s N_f\sim1$. In this framework, the $q\bar q$ bubble
corrections to the gluon propagator must be resummed to all orders.
Using  this dressed gluon propagator
one can then compute all the corrections with highest power of  $N_f$  at every $\alpha_s$ order to  a given QCD
observable~\cite{Renormalons}.
The results in large-$\beta_0$ are obtained by replacing the $N_f$
dependence by the leading QCD $\beta$-function coefficient ($\beta_1$ in our notation) which incorporates a set of non-abelian gluon-loop diagrams.
 Accordingly,
 the QCD $\beta$-function is truncated at its first term.\footnote{Strictly speaking,  the large-$\beta_0$ limit would  be the ``large-$\beta_1$'' limit, in our notation. }

 In this limit, the Borel transform of the reduced Adler function, defined in Eq.~(\ref{BorelDef}) can be written in a closed form  as~\cite{MB93,DB93,Renormalons}
\beq
B [\widehat D](u) = \frac{32}{3 \pi} \frac{e^{(C+5/3) u}}{(2-u)} \sum\limits_{k=2}^\infty{\frac{(-1)^k k}{[k^2-(1-u)^2]^2}},\label{BDLb}
\eeq
where the scheme parameter $C$ measures the departure from the $\MSb$, which
corresponds to the choice $C=0$.\footnote{The use of $C+5/3$ to
  parametrize the scheme is different from the conventions of
  Refs.~\cite{Renormalons,BJ08,BBJ13} but makes it easier to make
  contact with Ref.~\cite{BJM16}.} In the conventions of
Eqs.(\ref{eq:DCIPT}) and (\ref{BorelDef}) we have  $u = \beta_{1} t/(2\pi)$ with $\beta_{1}=\frac{9}{2}$ (for
$N_f=3$). The result clearly exhibits the renormalon poles, both the
IR, that lie along the positive real axis, and the UV ones, that
appear on the negative real axis. They are all double poles, with the
sole exception of the leading IR pole at $u=2$, related to the gluon
condensate, which is a simple one. This Borel transformed Adler function is
a meromorphic function but it is important for the subsequent
discussion to note that it is not of the Stieltjes type as can be proved by the computation of the 
determinantal necessary-conditions for a function to be of Stieltjes type~\cite{Pades} --- and can be
easily seen by the alternating sign of the different renormalon contributions. This already
anticipates the presence of complex-conjugated poles in our approximants.

  The Borel transform of the FOPT correction to the decay rate, $B[\delta^{(0)}]$, can be obtained inserting Eq.~(\ref{BDLb}) in the expression of Eq.~(\ref{eq:delta0}).  The contour integral can be done using the one-loop logarithmic running of $\alpha_s$ to give\cite{BJ08}\footnote{For details about the calculation, see also a related discussion in Sec. IV.B of Ref.~\cite{DVs2018}.}
\beq
B[\delta^{(0)}](u) = \frac{12}{(1-u)(3-u)(4-u)}\frac{\sin (\pi u)}{\pi u}B [\widehat D](u).\label{BorelDelta0}
\eeq
The analytic struture of this last Borel transform is much simpler
than that of $B[\widehat{D}](u)$. Now all the UV poles are simple
poles, because of the zeros of $\sin(\pi u)$.  For the same reason,
the leading IR pole of $B[\widehat D](u)$, at $u=2$, which is simple
in large-$\beta_0$, is cancelled in $B[\delta^{(0)}](u)$ --- a result
first pointed out in Ref.~\cite{BY92} for the Borel transformed spectral function. Our analysis with PAs benefits
greatly from these cancellations since the Borel transformed function
is now much less singular.\footnote{The fact that the only poles that remain double in Eq.~(\ref{BorelDelta0}) are the ones at $u=3$ and $u=4$ is not a coincidence. This reflects the fact that $\delta^{(0)}$ is maximally sensitive to the dimension-six and dimension-eight OPE condensates. This may have consequences  for the choice of weight functions employed in $\alpha_s$ analyses
   from $\tau$ decays that will be investigated elsewhere~\cite{inprep}.
} A simpler analytic structure can be much
more easily mimicked by the PAs. We also note that the leading UV pole
has a residue about ten times smaller than in the Adler function
counterpart. This, together with an enhancement of the residue of the double
pole at $u=3$, postpones the sign alternation of the series and enlarges the 
range of convergence of the Taylor series. PAs constructed to the expansion
of Eq.~(\ref{BorelDelta0}) benefit from these features of $B[\delta^{(0)}](u)$
and lead to smaller  errors by virtue of Pommerenke's theorem, granting
better coefficient's determination~\cite{Pades}.

The coefficients $c_{n,1}$ of the reduced Adler function can be reconstructed from the Borel transform by performing the expansion around $u=0$ and using  Eqs.~(\ref{RedAdler}) and~(\ref{BorelDef}). The first six coefficients of the Adler function in the large-$\beta_0$ limit, denoted $\wh D_{L\beta}$, read ($N_f=3$, $\MSb$)
\beq
\wh D_{L\beta}(a_Q) = a_Q + 1.556 \, a_Q^2 + 15.71\,  a_Q^3 + 24.83\,  a_Q^4 + 787.8\,  a_Q^5 -1991  \, a_Q^6+ \cdots, \label{DLb}
\eeq
to be compared with their QCD counterparts given in Eq.~(\ref{DinQCD}).  We observe that the sign alternation due to leading UV renormalon sets in at the sixth order (in the $\MSb$).
These coefficients lead to the following large-$\beta_0$ FOPT expansion of $\delta^{(0)}$:
\beq
\delta^{(0)}_{{\rm FO},L\beta}(a_Q) = a_Q + 5.119\,  a_Q^2 + 28.78\,  a_Q^3 + 156.7\,  a_Q^4 + 900.8\,  a_Q^5 +  4867 \, a_Q^6\cdots,\label{FOPTLb}
 \eeq
to be compared with Eq.~(\ref{FOPTQCD}). Now the sign alternation of
the coefficients is postponed and sets in only at the 9th order because
of the suppression of the leading UV pole in Eq.~(\ref{BorelDelta0}).
In comparison with the results in full QCD, the large-$\beta_0$ limit is
a good approximation, in the case of the Adler function, only up to
$\alpha_s^2$. However, for   $\delta^{(0)}_{{\rm FO},L\beta}$ this
approximation is still good up to the last known term,
i.e. $\alpha_s^4$. The reason for this better agreement lies in the
fact that these coefficients depend also on the $\beta$-function
coefficients --- which are largely dominated by $\beta_1$ in QCD ---
as well as on the integrals of Eq.~(\ref{FOPTdef}).

An important difference between the $a_Q$ expansion of the Adler
function and that of $\delta^{(0)}$, Eqs.~(\ref{DLb})
and~(\ref{FOPTLb}) is that, in the former, the smallest term of the
sum is reached already at the fourth order. This makes the asymptotic
nature of the series very prominent. In FOPT, the series is much
better behaved and each term is consistently smaller than the previous
up to the 9th order.  The FOPT series, therefore, behaves at
intermediate orders almost as a convergent series --- a fact that will
be important in the remainder of the paper. (These features can be
visualized in the results represented by solid lines in
Figs.~\ref{P12MSbLbAdler} and~\ref{P12MSbLbdelta}.) It can be shown
analytically that in the coefficients $d_n$ of Eq.~(\ref{FOPTexp})
there are cancelations between the Adler function coefficients and the
remainder contributions~\cite{BJ08}.  These cancellations lead to the fact that  FOPT is,
in large-$\beta_0$,  superior to CIPT (which misses them). Aditionally, the cancellations suppress to some extent
the divergent character of the series, which is postponed with respect to the
Adler function.

A special feature of the large-$\beta_0$ result is the simple way in
which the scheme dependence appears through the factor $e^{(C+5/3)u}$ in Eq.~(\ref{BDLb}).  It becomes clear that the residue
of the renormalon poles is scheme dependent, while their position,
related to the dimension of operators in the OPE, is not. Physical
results must, of course, be scheme independent.  However, the coupling
$\alpha_s$ is itself not physical, since it depends on conventions
related to the renormalization procedure. Therefore, a perturbative
expansion in $\alpha_s$ is a scheme-dependent approximation to an
(unknown) scheme-independent physical result.

In this context, the physical result is given by the Borel integral
Eq.~(\ref{BorelInt}) in which the scheme dependence of the Borel
transform is cancelled by the scheme dependence of the coupling
$\alpha_s$, denoted by $\alpha_s^C$, to make it explicit.  Writing the Borel transform as
\beq
B[\widehat D] (u) = e^{C u} b(u),
\eeq
the function $b(u)$ is scheme independent and we have
\beq
\widehat D (\alpha) \equiv \int\limits_{0}^{\infty} 
dt \exp\left[-t\left(\frac{1}{\alpha^C_s}- \frac{\beta_1 C}{2\pi}   \right) \right] 
b\left(\frac{\beta_1 t}{2\pi}\right),
\label{BorelIntLb}
\eeq
which exposes the scheme invariant combination
\beq
\frac{1}{\bar \alpha} = \frac{1}{\alpha_s^C} - \frac{\beta_1 C}{2\pi}.
\eeq
This result  allows us to write the coupling $\alpha_s^C\equiv \hat\alpha$ in
terms of the more usual $\MSb$ coupling as
\beq
\frac{1}{\alpha_s^C} \equiv\frac{1}{\hat \alpha_s}= \frac{1}{\alpha_s^{\MSb}} + \frac{\beta_1 C}{2\pi} . \label{alphahat}
\eeq
Redefinitions of the QCD coupling of this type have been discussed in
Refs.~\cite{BJM16,JM16}. The result we employ here is a particular case of
those of Ref.~\cite{BJM16}, with higher order $\beta$-function
coefficients set to zero. Since the QCD $\beta$-function is dominated by
$\beta_1$, the qualitative behavior of $\hat \alpha_s$ with $C$
remains the same as in Ref.~\cite{BJM16}: {\it grosso modo}, negative values correspond to larger $\hat \alpha_s$  whereas positive values of $C$ are
associated with smaller $\hat \alpha_s$ values.

Here, we will  exploit the freedom of scheme choice in order to
optimize the rational approximation of the Borel transformed Adler
function. It is particularly important to note that the schemes with
negative $C$ values, therefore less perturbative, introduce a
suppression of the IR pole residues. In these schemes, IR pole
contributions are largely dominated by the first few poles, much more so
than in schemes with $C>0$. On the other hand, for $C<0$ the UV poles are
enhanced, and one expect the sign alternation of the series to show up
at very low orders. For perturbative calculations, these schemes with larger
 $\hat \alpha_s$ values are essentially useless, but we will show that they are more amenable to a
rational approximation as $C<0$ suppresses the influence of the
exponential term in Eq.~(\ref{BDLb}) and results in a function with
more pronounced isolated poles, easier to reproduce with a rational
function than an exponential one.


\subsection{Pad\'e approximants to the Adler function}
\subsubsection[$\MSb$ scheme] {\boldmath $\MSb$ scheme}
\label{MSbLb}

We begin by using Pad\'e approximants to study the perturbative
expansion of the Adler function and of $\delta^{(0)}$ in the $\MSb$
scheme.  Since in large-$\beta_0$ we know the exact result we are able
to assess the quality of the approximation and refine the method that
later we will apply to QCD. In the remainder of this section, we
devise a strategy to extract as much information as possible about the
series using rational approximants. 

Let us first comment on the construction of Pad\'e
  approximants directly to the series in $\alpha_s/\pi$, given of
  Eq.~(\ref{DLb}).   In the case of the Adler
  function, the asymptotic nature of the series is very prominent since  
  from the 5th term on asymptoticity has already set
  in. Forming Pad\'e approximants to the Adler series in $a_Q$  
   requires many coefficients as input in order to
  allow for an acceptable description of higher orders. The Borel
  transformed Adler function, which suppresses the factorial growth of
  the coefficients fixes, at least partially, this behaviour and is much
  more amenable to the approximation by PAs. This has been noted
  already in Ref.~\cite{SEK95} and we refrain from  further discussing PAs 
  constructed for the  $\alpha_s/\pi$ expansion of the Adler function. (PAs of this type will turn out to be
  useful, however, in the case of $\delta^{(0)}$, as discussed in Sec.~\ref{PAsToDelta0})

We turn now to Pad\'e approximants formed to the Borel transformed Adler function.
The first question that arises regards what Pad\'e sequence(s) to use.
In our conventions, the $\MSb$ corresponds to $C=0$, which means that
the exact Borel transform diverges exponentially when $u\to
\infty$. Since we do not have a simple power-like behaviour  for
large $u$ we do not attempt to fix the Pad\'e sequence using this
limit. Instead, we will
investigate more than one sequence, keeping in mind that in QCD we
have only the first four coefficients of the Adler function available.

Let us begin with the sequence $P^{N+1}_N(u)$. Since we are
interested in the prediction of the behaviour of the series at higher
orders, we start by studying the quality of the estimate of the first
coefficient not used as input. Each $P^{N+1}_N$ needs $2N+2$
parameters to be constructed and we employ the first $2N+2$
coefficients of the Borel transformed Adler function. 
Then the Pad\'e is reexpanded to predict the coefficient $c_{2N+3,1}$.
In the first non trivial  case, that of $P_1^2(u)$, we need $c_{1,1}$, $c_{2,1}$, $c_{3,1}$, and $c_{4,1}$ to fix the parameters and  we find
\beq
P^2_1 (u) = \frac{1.359 + 0.6221 u + 1.889 u^2}{ 4.271-u},\label{P12MSbLb}
\eeq
from which we extract the value $52.33$ for $c_{5,1}$ to be compared
with the exact value 787.8, given in Eq.~(\ref{DLb}) --- clearly not
an accurate prediction.  For the FOPT series this leads to the value
165.3 for the fifth coefficient, to be compared with 900.8 in
Eq.~(\ref{FOPTLb}). The approximant $P_1^2(u)$ displays an effective pole at 4.271,
which cannot be straightforwardly associated with any of the poles of
the exact Borel transform. This is the aforementioned feature of low-order Pad\'e approximants:
they mimic the infinite tower of poles by the appearance of effective poles not present in the original function.
In a Pad\'e with several poles, only the first few, closest to the origin, can be identified
with the poles of the original function and only in a hierarchical way~\cite{MP07}.\footnote{This poses a word
  of caution in the interpretation of results from renormalon models with a small number of fixed poles. In these cases,
  the only freedom left is in the residues that must accommodate the imperfections of the model in an effective way. Therefore,
  for the same reasons discussed above, residues of poles further away from the origin may not correspond to their counterpart in
the original function.}

\begin{figure}[!t]
\begin{center}
\includegraphics[width=.65\columnwidth,angle=0]{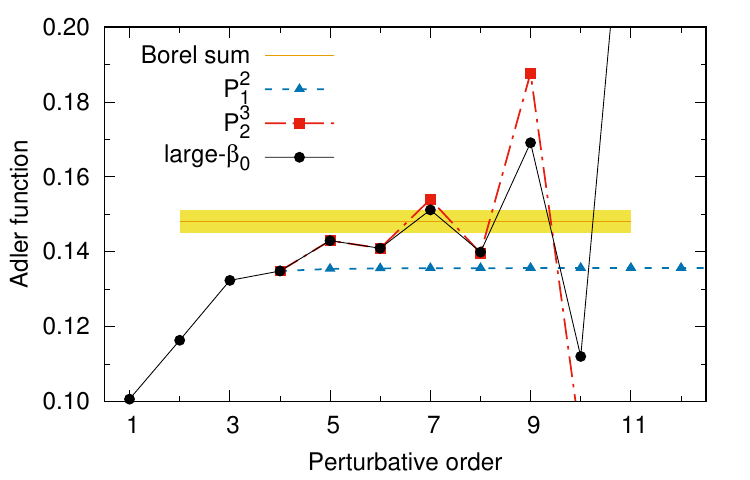}
\caption{Results for the perturbative expansion of the Adler function in large-$\beta_0$ (solid line) and using the $P_1^2(u)$ (dashed line) and $P_2^3(u)$ (dot dashed line). The result of the Borel sum of the series is displayed with a band that represents its ambiguity. In all figures we use $\alpha_s(m_\tau^2)=0.316\pm 0.010$~\cite{PDG}.}
\label{P12MSbLbAdler}
\end{center}
\end{figure}

It is illustrative to consider the next Pad\'e approximant in this sequence, $P_2^3(u)$, and compare the results.
 In this case, we find
\beq
P^3_2 (u) = \frac{0.3385 + 0.4005 u + 0.3219 u^2 + 0.1609 u^3}{( 0.8024+u)(1.325 - u)}.
\eeq
Now the first coefficient that can be forecast is $c_{7,1}$ for which
we find the value 125,745, to be compared with the exact value
98,572.8.  The 7th order coefficient of the FOPT series is also much
better forecast: 59,456.6 to be compared with the exact coefficient
32,284.3. This approximant is able to reproduce the qualitative
behaviour of the Adler function and the FOPT series at higher orders,
mainly due to the fact that it exhibits a pole at $u=-0.8024$, which
mimics the leading UV pole at $u=-1$, and a second pole mimicking the
first IR pole, but slightly below the exact value at $u=2$. The UV
pole is enough to ensure the correct sign alternation in the higher
order coefficients, as shown in Fig.~\ref{P12MSbLbAdler} (throughout this
paper we use $\alpha_s(m_\tau) = 0.3160$, which corresponds to the most recent
PDG recommendation evolved to the $\tau$ mass scale~\cite{PDG}).

\begin{figure}[!t]
\begin{center}
\subfigure[$\delta^{(0)}$, FOPT.]{\includegraphics[width=.48\columnwidth,angle=0]{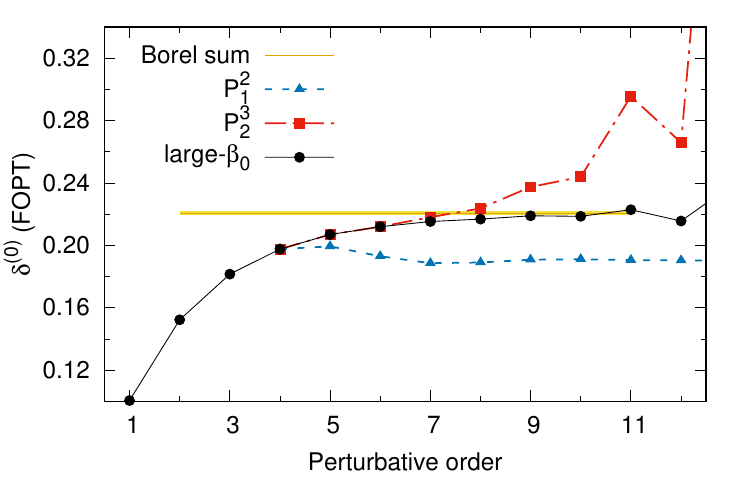}\label{P12MSbLbFOPT}}
\subfigure[$\delta^{(0)}$, CIPT.] {\includegraphics[width=.48\columnwidth,angle=0]{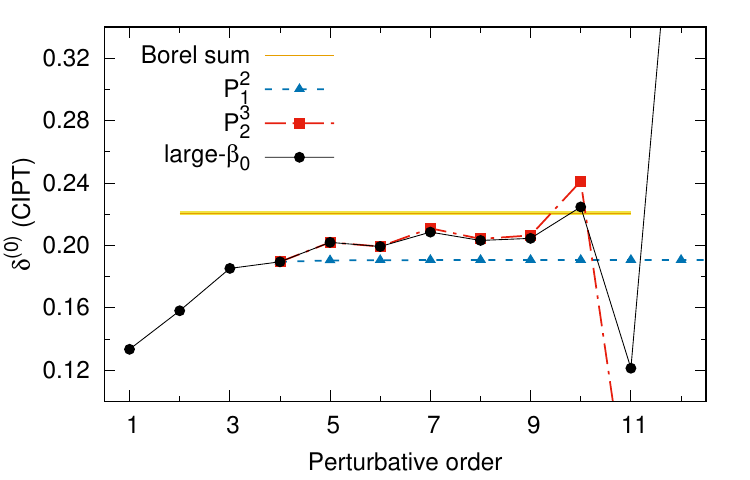}\label{P12MSbLbCIPT}}
\caption{Results for $\delta^{(0)}$ order by order in perturbation theory for large-$\beta_0$ (solid line) and for the  $P_1^2(u)$ (dashed line) and $P_2^3(u)$ (dot dashed line) in FOPT (left) and CIPT (right). The Borel sum of series with its ambiguity is shown as the horizontal band. }
\label{P12MSbLbdelta}
\end{center}
\end{figure}

A clear feature starts to emerge already at this level. To get an
appropriate approximation to the Adler function in the $\MSb$, at least two-pole approximants must be
considered. This provides the balance between UV and IR renormalon
contributions.  A visual account of the quality of these approximants
is given in Fig.~\ref{P12MSbLbAdler} which displays the exact Adler
function in large-$\beta_0$ and the result reconstructed from
$P_1^2(u)$ and $P_2^3(u)$, whereas the results for $\delta^{(0)}$ in
FOPT and CIPT are shown in Fig.~\ref{P12MSbLbdelta}.

Using the same amount of information, we could consider the
$P^N_{N+1}(u)$ sequence with the $P^1_2(u)$ and $P^2_3(u)$ its firsts
elements. As we observed before, two-pole approximants yields better
convergence ranges (by capturing the sign-alternating feature of the
Borel series) so we should expect better $c_{5,1}$ and $c_{7,1}$
predictions for this sequence. We actually find $c_{5,1} = 1770$ and
$c_{7,1}=102,889$ respectively, a better determination than their
$P_N^{N+1}(u)$ counterparts. 

It is thus interesting to investigate systematically the convergence of the Pad\'e
sequence with respect to $N$. In order to quantify the
quality of the prediction of  coefficients not used as input
we define the relative error as
\beq
\sigma_{\rm rel} = \left| \frac{ c_{n,1}^P-c_{n,1} }{c_{n,1}}\right|,\label{sigmarel}
\eeq
where $c^P_{n,1}$ is the coefficient extracted from the Pad\'e
approximant. If the PA sequence converges, the parameter $\sigma_{\rm rel}$
should tend to zero as $n$ grows.

\begin{figure}[!t]
\begin{center}
\subfigure[]{\includegraphics[width=.49\columnwidth,angle=0]{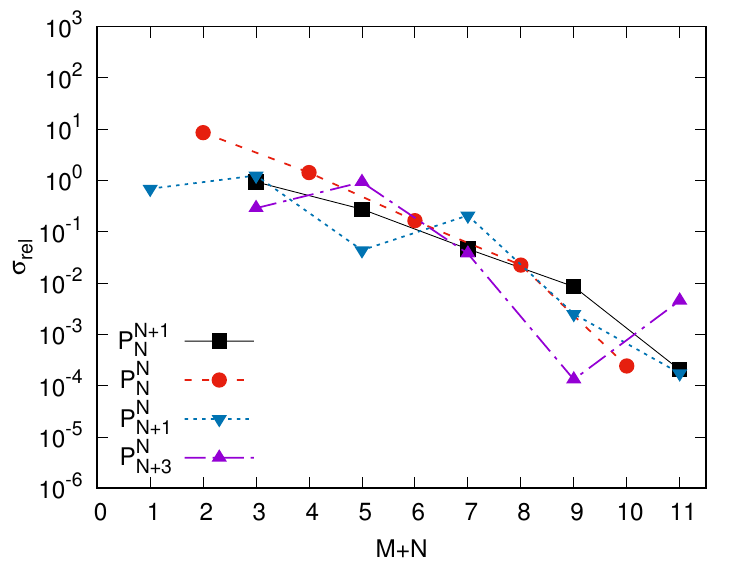}\label{Error_Lb_MSb}}
\subfigure[] {\includegraphics[width=.49\columnwidth,angle=0]{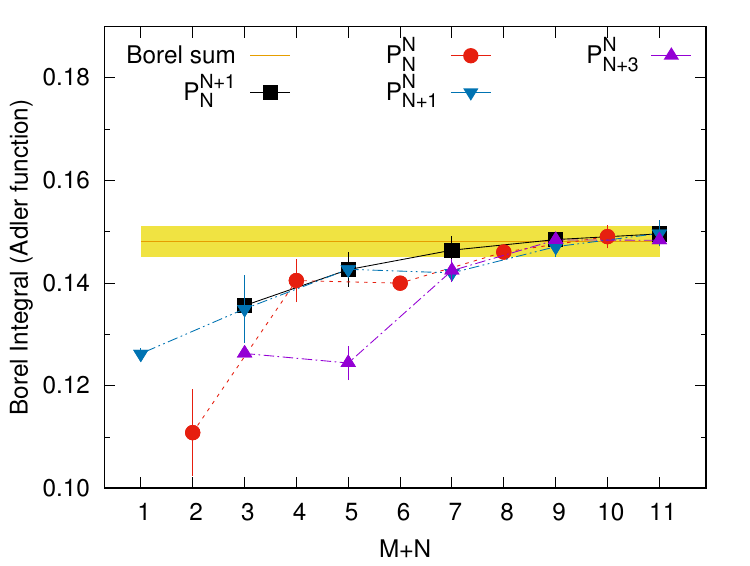}\label{BorelInt_Lb_MSb}}
\caption{Results for four different Pad\'e sequences in the $\MSb$ scheme.  (a) Relative error of the first forecast coefficient of the Adler function, as defined in Eq.~(\ref{sigmarel}). (b) Results for the Borel sum, Eq.~(\ref{BorelInt}). The horizontal band gives the imaginary ambiguity of the true value. 
}
\label{Fig3}
\end{center}
\end{figure}

In Fig.~\ref{Error_Lb_MSb} we show the behavior of
the relative error of the estimate of the $c_{2N+3,1}$ coefficient for
the sequence $P_N^{N+1}$ as a function of $N$. Being the Adler function in large-$\beta_0$
a meromorphic function, the Pommerenke's theorem ensures the convergence
for a larger set of PA sequences than the $P_N^{N+1}$. To show this excellent
global convergence pattern, we collect in Fig.~\ref{Error_Lb_MSb} few of the
closest-to-diagonal sequences, in particular the $P^{N+1}_N$, $P_N^N$, $P_{N+1}^N$, and $P_{N+3}^N$.

The results are qualitatively similar for all sequences: in all cases
the convergence to the exact results happens fast as $N$ grows.  We
observe in Fig.~\ref{Error_Lb_MSb} that the improvement once $N$ is
increased by one unity is often of about one order of magnitude or
more (notice the log scale of the plot).  There are a few outliers
where increasing $N$ by one unity does not represent an improvement,
or even makes the relative error larger. A close scrutiny of the
particular approximants reveals why this is so. For meromorphic
non-Stieltjes functions, it is a feature of the Pommerenke's
theorem~\cite{Pades,MP07} that the convergence pattern can be altered
by the presence of defects, transients poles almost cancelled by a
close-by zero as it is clearly noticed after observing the
convergence pattern of the pole positions for the $P^N_{N+1}$
sequence: $P^1_2(u)$ has poles at $u=-0.56$ and at $u=0.89$;
$P^2_3(u)$ at $u=-0.86$ together with a couple of complex-conjugated
(CC) poles at $u=1.43 \pm 0.46 i$ --- notice the stability of the UV
pole which is driving the large-order behavior of the series (since it
is always the closest to the origin)--- then, $P^3_4(u)$ contains
poles at $u=-0.85$, the CC poles at $u= 1.40 \pm 0.47 i$ and an
extraneous new pole closer to the origin at $u=-0.3991$.
This last one, which would
eventually spoil the convergence, is a new pole and it is actually
canceled by a close-by zero at $u=0.3989$  effectively reducing the order of this approximant to a $P^2_3$.   A similar feature is observed for the
second and fifth elements of the $P^N_{N+3}$ sequence. In this last
case, the cancellation among zero and pole is of the order of
$10^{-13}$, i.e., the residue of the spurious pole is ${\cal  O}(10^{-13})$. Identifying these cancelations will be important in the
obtention of the final results of this paper.

When $N=4$, in all sequences the results are better by a few orders of magnitude
compared to the first PA in each sequence. We remark that the
convergence of some of these sequences was already studied in
Ref.~\cite{SEK95} and we confirm and extend their results.

In addition, PAs can be used as a way to resum the
original series. In our case, the Borel integral can be performed with
the reconstructed PAs using Eq.~(\ref{BorelInt}) in order to produce an
estimate for the Borel sum of the series.
The results for the Borel sum of the series also approach the true
value when the order $N$ is increased, as can be seen in
Fig.~\ref{BorelInt_Lb_MSb}. As expected, the approximation becomes
increasingly better, but it is also worth noting that the error is
significantly reduced only after a relatively large value of $N$ is
employed.

All in all, we arrive to the observation that PAs with larger values
of $N$ tend to have poles that can be identified with the leading
renormalons --- at least for the ones that are closer to the origin in
a hierarchical way~\cite{MP07}--- and this is enough to yield a good
approximation to the series coefficients and the Borel integral. In
all cases, however, realistic values of $N$ ($N=1$ for $P_N^{N+1}$ and
$P_{N+1}^N$ and $N=0$ for $P_{N+3}^N$) still do not provide a good
approximation, as can be seen in Figs.~\ref{P12MSbLbAdler},
\ref{P12MSbLbdelta}, \ref{Error_Lb_MSb} and \ref{BorelInt_Lb_MSb}.
Since in full QCD only the first four coefficients of the Adler
function are known it would be desirable to obtain a better
approximation for these lower values of $N$. In the next section we
discuss how scheme variations can be used to that end.


\subsubsection{Scheme variations}

The exact result for the Borel transformed Adler function in
large-$\beta_0$, Eq.~(\ref{BDLb}), displays explicitly an important
property: the residues of the renormalon poles are scheme dependent
but their position is not. Here we exploit this feature in order to
improve upon the results of the previous section.

One of the difficulties in using PAs with a small number of parameters
is the fact that they must mimic an infinite tower of renormalon poles
and their complicated interplay with a set of only a few  poles. In the
$\MSb$, in lower orders, the first UV and IR poles give sizeable
contributions to the coefficients~\cite{BJ08,BBJ13}. However, using
schemes with negative values of $C$ the Borel transform becomes much
more dominated by the first UV pole, due to the factor of
$e^{(C+5/3)u}$, and the sign alternation should show up at very low
orders. Because of this dominance, it becomes easier for a PA to
reproduce such a series (we will elaborate on that below). Of course,
negative values of $C$ entail larger values of $\hat
\alpha_s$~\cite{BJM16}, as per Eq.~(\ref{alphahat}). These schemes are
therefore bad for perturbative calculations but they are very useful,
for example, to obtain estimates for the Borel sum of the series,
since this is a scheme-independent result. Finally, the results for
the coefficients with negative $C$ values can be translated to the
$\MSb$ using the relation between the two couplings, and we will show
that this leads to better results for the predicted higher order coefficients
in $\MSb$.

\begin{figure}[!t]
\begin{center}
\includegraphics[width=.65\columnwidth,angle=0]{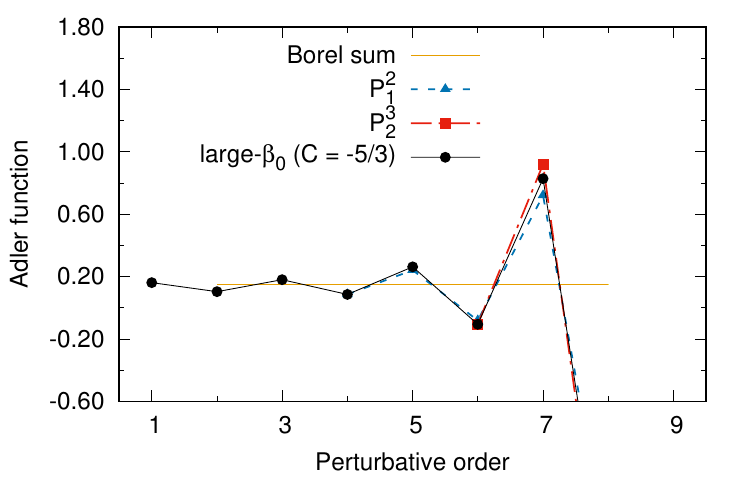}
\caption{Perturbative expansion of the Adler function in large-$\beta_0$ in a scheme with $C=-5/3$ (solid line)  and using the $P_1^2(u)$ (dashed line) and $P_2^3(u)$ (dot dashed line). The scheme-invariant result for the Borel sum of the series is the same as in Fig.~\ref{P12MSbLbAdler}. The value of the strong coupling in this scheme is $ \hat \alpha_s (m_\tau) = 0.5074$ (see text).}
\label{Adler_Cm2_Lb}
\end{center}
\end{figure}

For definiteness, we will work with $C=-5/3$ which cancels exactly the exponential in Eq.~(\ref{BDLb}). In this scheme the central value of the coupling is $ \hat \alpha_s (m_\tau) = 0.5074$.  The expansion of the Adler function in the large-$\beta_0$ limit for $C=-5/3$ reads ($N_f=3$)
\beq
\wh D_{L\beta}^{(C=-5/3)}(a_Q) = \hat a_Q  -2.194 \, \hat a_Q^2 +18.10\,  \hat a_Q^3  -139.0\,  \hat a_Q^4 + 1610\,  \hat a_Q^5 - 20,759\,\hat a_Q^6 + \cdots \label{DCm2}
\eeq
We remark that, as expected, the sign alternation sets in much
earlier, in this case it starts from the second coefficient. The exact
result for the Adler function in the large-$\beta_0$ limit for $C=-5/3$
can be seen in Fig.~\ref{Adler_Cm2_Lb}. It becomes clear that asymptoticity sets in also
earlier due to the much larger value of the expansion parameter, $\hat
\alpha_s$.

Let us consider again the Pad\'e $P_1^2(u)$, but now in the scheme
$C=-5/3$. In this new scheme we find
\beq
P_1^2 (u) =\frac{0.2798+0.0455\, u +0.1899\, u^2}{0.8790+u}, \qquad
(C=-5/3). \label{P12Cm2Lb}
\eeq
Now, for the coefficient $c_{5,1}$ the
result is $1423$, to be compared with the exact value $1610$ in
Eq.~(\ref{DCm2}), a relative error of only $12 \%$ --- this is an
improvement of about an order of magnitude with respect to the result
in the $\MSb$. The result for $c_{6,1}$, one order higher, is still
very good: $-18,212$, just a $12\%$ relative error.
Also, this $P_1^2(u)$ already displays a pole at $-0.8790$, close to the leading UV
renormalon, and which reproduces the sign alternation of the
coefficients. Finally, the Borel integral gives now 0.1416, much closer to the exact value (which is $0.1481\pm 0.0030i$, shown in Fig.~\ref{BorelInt_Lb_MSb}).  In Fig.~\ref{Adler_Cm2_Lb}, we see that the agreement with
the exact Adler function is  good even at higher orders.  
Clearly,  $P_1^2(u)$ in this scheme provides a much more accurate prediction of the
true function than its counterpart in the $\MSb$. The underlying reason is simple: without the exponential term,
the position of the most prominent renormalon pole, the first UV, is much better determined in full agreement
with the Pommerenke's theorem~\cite{Pades,MP07}. Furthermore, the suppression
of the IR-pole residues simplify the interplay of the poles. Enlarging the sequence will only improve on the results.

Actually, for $C=-5/3$, the improvement for the $P_2^3(u)$ is only modest. We find
\beq
P_2^3(u) = \frac{0.2299-0.2080\,u+0.0769\, u^2 -0.1283\,u^3}{(0.8757-u)(0.8248+u)}   ,\qquad (C=-5/3)\, ,
\eeq
whit a prediction of the 7th coefficient has now a relative error of $10\%$ and pretty stable position of the pole closest to the origin: $-0.8248$.
 The Borel integral is again well predicted: $0.1394 + 0.0048i$. 

For the next element in the sequence,  $P_3^4(u)$, the prediction of the 9th coefficient has an error of mere $0.19\%$, about two orders of magnitude better than in the
$\MSb$. Interestingly, for the first time, this PA has two poles close to $u=-1$, one
at $-1.1312$ and a second at $-0.9385$ which mimics the fact that the
leading UV pole is, actually, a double pole. It also has a pole at
$1.788$, rather close to the location of the first IR pole, which lies
at $u=2$.  Finally, its Borel integral is also almost on top of the true one:
$0.1476 \pm 0.0084\,i$, the real part is off by only 0.3\%.

\begin{figure}[!t]
\begin{center}
\subfigure[]{\includegraphics[width=.49\columnwidth,angle=0]{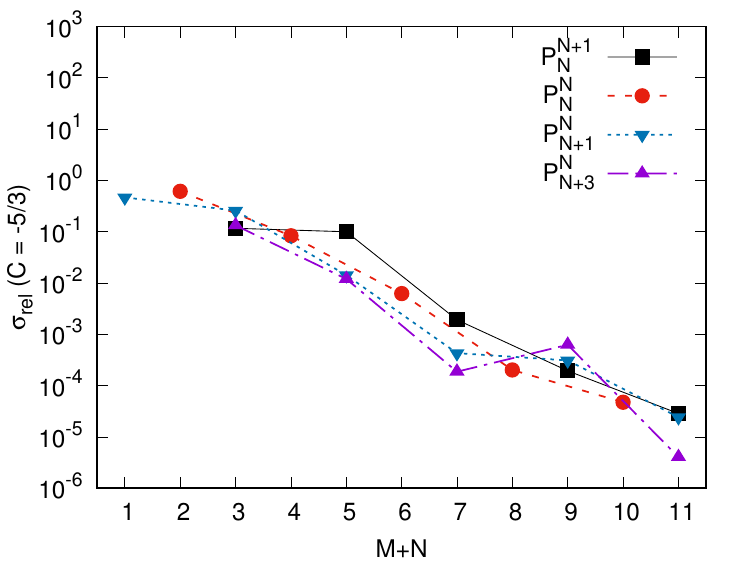}\label{Error_Lb_Cm2}}
\subfigure[] {\includegraphics[width=.49\columnwidth,angle=0]{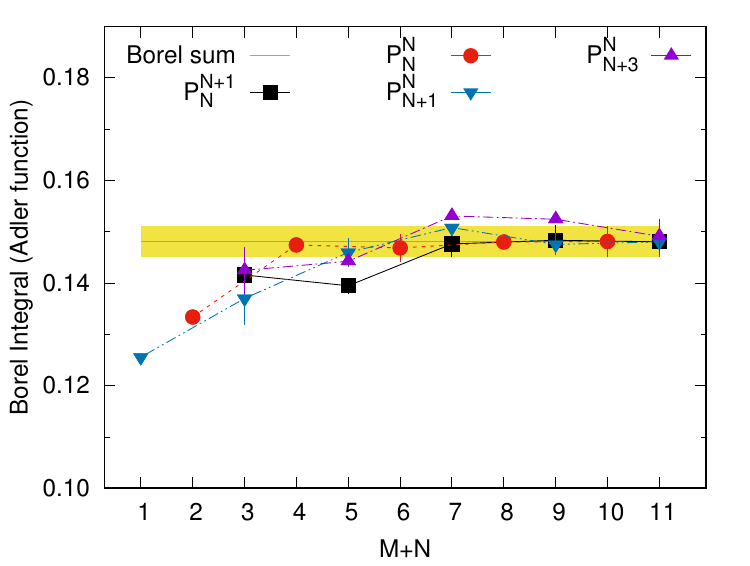}\label{BorelInt_Lb_Cm2}}
\caption{Results for different Pad\'e sequences in a scheme with $C=-5/3$.  The  scales are the same as in Fig.~\ref{Fig3} to facilitate the comparison. (a) Relative error of the first forecast coefficient of the Adler function. (b) Results for the Borel sum, Eq.~(\ref{BorelInt}). }
\end{center}
\end{figure}

Again, this excellent convergence pattern is not particular for this
sequence.  In Fig.~\ref{Error_Lb_Cm2} we show the relative error of
the first predicted coefficient using the same four sequences of
Fig.~\ref{Error_Lb_MSb}. The comparison with the results in the $\MSb$
clearly shows the advantage of using less perturbative schemes. For
instance, the results for $P_N^{N+1}$ for $N=1$ are as good as those
of the $\MSb$ with $N=3$, but require four parameters less. The
results for the scheme-independent Borel sum predicted by these Pad\'e
sequences follow suit and are also significantly better than in the
$\MSb$, as Fig.~\ref{BorelInt_Lb_Cm2} shows. We should remark that for
most of the Pad\'es the results are even better if $C$ is lowered
below $-5/3$.\footnote{For $P_1^2$, e.g., $C=-2$ provides an
  excellent description of the Adler function up to order 9 with only
  4 parameters. However, other negative values of $C$ are somewhat
  arbitrary and the results must be checked for
  stability.  }  This
means that the improvement in the results should not be directly
attributed to the cancellation of the exponential term but rather to
the value of the argument of that exponential.  It is well-known~\cite{Pades}
that the pole
positions of diagonal and near-diagonal PA to an exponential function
share three characteristics: the location corresponds with the sign of
the exponential argument (positive argument, positive location in the
complex plane with positive real part, and vice-versa), as soon as the
PA order increases, the poles move further away from the origin, and
finally for a given PA, the position of its poles is located within an
area defined by its ``order star'' (cf. Ref.~\cite{Pades} for further
definitions).  These implies that as soon as the scheme is less
perturbative, the PA poles responsible for the exponential term accumulate
in the UV region further away  as the PA order increases,
isolating the rest of the poles coming from the non-exponential term,
in particular the dominant UV double pole. On the contrary, for more
perturbative schemes, PA poles will accumulate in the IR region and
shadow the UV pole.  From this perspective it is then natural to
expect better convergence with respect to series coefficients and the
Borel integral for the less perturbative schemes.  It is apparent as well
that the $C$-scheme analysis sheds light on the analytical structure of
the Adler function in a clear way.

Let us now translate the results obtained in the scheme with $C=-5/3$ to the $\MSb$. The
expansion of Eq.~(\ref{alphahat}) gives the following perturbative relation
between the different schemes
\beq
\hat a_Q = a_Q + \frac{C \beta_1}{2} a_Q^2 + \Big{(} \frac{C \beta_1}{2}\Big{)}^2 a_Q^3 + \Big{(}\frac{C \beta_1}{2}\Big{)}^3 a_Q^4 +\cdots, \label{schemerelation}
\eeq
where on the r.h.s we have $a_Q$ in the $\MSb$ and on the l.h.s. $\hat
a_Q\equiv a_Q^{C}$. Applying this perturbative relation to an
expansion such as Eq.~(\ref{DCm2}) one is able to reconstruct the
$\MSb$ result from its counterpart in a scheme with a different $C$
value. Performing this procedure for the results of $P_1^2(u)$, $P_2^1(u)$ and $P_2^3(u)$
constructed at $C=-5/3$ leads to much better predictions of the higher
order coefficients, as Tab.~\ref{LbCoeff} confirms. The predictions
are far superior to those obtained when the PA is constructed directly in
the $\MSb$.

\bgroup
\def\arraystretch{1.3}%
\begin{table}[!t]
\begin{center}
\caption{Coefficients of the Adler function in the large-$\beta_0$ approximation in the $\MSb$ scheme. The first row gives the exact values. Darker rows show the results obtained from $P_1^2(u)$, $P_2^1(u)$ and $P_2^3(u)$ constructed in a scheme with $C=-5/3$ and later evolved to $\MSb$ using Eq.~(\ref{schemerelation}). For comparison, we also display the results obtained through the use of these PAs constructed directly in the $\MSb$ (second, fourth and sixth rows).}
\begin{tabular}{lcccccc}
\toprule
& $c_{5,1}$ & $c_{6,1}$ & $c_{7,1}$ &   $c_{8,1}$ & $c_{9,1}$  \\
 \midrule
 large-$\beta_0$ exact &   $787.8 $      & $-1991$    &   $9.857\times 10^{4}$  &  $-1.078\times10^6$  &  $2.775\times10^7$      \\
 $P_1^{2}$ ($\MSb$)      &   $52.33$        & $137.9$    &   $4.358\times 10^2$   & $1.605\times 10^{3}$  & $2.186\times 10^{4}$    \\
\rowcolor{light-gray} $P_1^{2}$ ($C=-5/3$)    &    $600.5$      & $-2958.0$    &   $7.022\times 10^{4}$  & $-1.134\times10^{6}$  & $2.382\times 10^{7}$    \\
 $P_2^{1}$ ($\MSb$)      &   $1770.1$        & $-8123.9$    &   $61.277\times 10^4$   & $-9.857\times 10^{6}$  & $4.665\times 10^{8}$    \\
\rowcolor{light-gray} $P_2^{1}$ ($C=-5/3$)    &    $1205.3$      & $-1722.4$    &   $26.980\times 10^{4}$  & $-2.024\times10^{6}$  & $1.234\times 10^{8}$    \\
$P_2^{3}$  ($\MSb$)     &     input       &  input      &  $1.257\times 10^{5}$   &$-1.372\times 10^{6}$ & $4.566\times 10^{7}$     \\
\rowcolor{light-gray} $P_2^{3}$   ($C=-5/3$)   &     input       &   input     &   $1.311\times 10^{5}$ & $-9.721\times 10^{5}$& $5.123\times 10^{7}$      \\
 \bottomrule
\end{tabular}
\label{LbCoeff}
\end{center}
\end{table}
\egroup


\subsubsection{Partial Pad\'e Approximants}

We have seen that Pad\'e sequences appear to converge rather fast to
the Borel transform of the Adler function, as expected following
Pommerenke's convergence theorem.  In realistic applications, however,
where one has only the first four coefficients of the series it may be
necessary to employ a method to accelerate the convergence such as the
scheme variation we discussed previously.  In this section we will
show that using knowledge about the position of the renormalon
singularities  significantly accelerates the convergence
and yields excellent results even for the lowest approximants.
We then consider now partial Pad\'e approximants (PPAs) as defined in
Eq.~(\ref{PartialPades}). 

Most of the effort done by the PAs in the previous sections was to
locate the double pole at $u=-1$, and this had a cost of two series
coefficients. It is then to expect that including the double pole in
advanced should allow the approximants to unfold the subdominant
renormalon poles. For the $\MSb$ scheme, a $\mathbb{P}^{2}_{1,2}(u)$
(imposing the double pole at $u=-1$), leads to a prediction of
$c_{5,1}$ which is $60\%$ off, while $c_{5,1}$ from
$\mathbb{P}^{1}_{2,2}(u)$ is $55\%$ off. This represents a significant
improvement with respect to the PAs. Results are much better for
$C=-5/3$ where the predictions reach precision better than $3\%$ for
both. In both schemes, pole predictions suggest the subdominant
renormalons to be located in the IR region. Using more series
coefficients one identifies the $u=2$ as the first IR renormalon.

As we have seen, the Borel transform of the Adler function has indeed
an IR single pole at $u=2$ and the next step towards improving
precision would be to consider a $\mathbb{P}^{M}_{N,3}(u)$ including
such pole.  In this case, and for the $\MSb$ scheme, results 
improve since with a $\mathbb{P}^{2}_{1,3}(u)$ one gets $30\%$ of
relative error on the $c_{5,1}$ determination, whereas  $20\%$ is reached
with the $\mathbb{P}^{1}_{2,3}(u)$. For the  $C=-5/3$ scheme, the
result is greatly improved since relative errors reach up to $1\%$ and
$5\%$ respectively. In this scheme, even a Pad\'e-type approximant
with a fixed denominator at $(u+1)^2(u-2)$, a
$\mathbb{P}^{M}_{0,3}(u)$, yields  good results.  In this sort of
sequence, one can even study the convergence by looking systematically
at the $c_{5,1}$ prediction for growing $M$. We find $35\%$, $7\%$,
and $4\%$ relative error for $M=1,2,3$ respectively.

Unraveling the sub-subdominant renormalons  is now more an art than a science since it is difficult to decide whether
the second UV or the second IR should be considered. The decision comes from exploring systematically the
residue of the predicted poles from previous approximants. 
We observe that they predict an IR (double) pole at around $u=3$ with large residue. As such, it contributes
largely to the series coefficients. The next step will be then to consider $\mathbb{P}^{M}_{N,5}(u)$ including this double pole where the polynomial $T_5(u)$ of Eq.~(\ref{PartialPades}) is constructed such as
to reproduce the first five poles of the Borel transformed Adler
function
\beq
T_5(u) = (u+1)^2(u-2)(u-3)^2.
\label{T5}
\eeq
We can then study near diagonal sequences akin to the ones we discussed
in the previous section, e.g., $\mathbb{P}_{N,5}^{N+1}(u)$,  $\mathbb{P}_{N+1,5}^{N}(u)$, or  $\mathbb{P}_{N,5}^{N}(u)$. 
It is expected that these sequences should yield much better results since  the perturbative series is dominated at intermediate and large orders by the poles closest to the origin.

We start with results for the sequence $\mathbb{P}_{N,5}^{N+1}(u)$
constructing the PPAs in the $\MSb$ scheme.
The relative error of the the coefficient $c_{5,1}$
obtained from $\mathbb{P}_{1,5}^2(u)$ is 12 times smaller than the one
from $P_{1}^2(u)$.  The four coefficients used to fix the parameters of
$\mathbb{P}_{1,5}^{2}(u)$ provide enough information to obtain a good
approximation even after asymptoticity has set in --- the 10th
coefficient is predicted with a relative error of only $20\%$. This
reproduction of the series is achieved by generating a pole at 1.750. 
Results from $\mathbb{P}_{2,5}^{3}(u)$ are so similar to the
original function that they cannot   be distinguished by eye. The
approximant $\mathbb{P}_{2,5}^{3}$ has a pair of complex poles at
$2.854\pm 0.992 \,i$ which, again, appear due to the fact that the meromorphic
function being approximated is not of the Stieltjes type.

The use of the scheme with $C=-5/3 $ accelerates the convergence and
now the results from $\mathbb{P}_{1,5}^{2}(u)$ cannot be visually
distinguished from the exact ones (we therefore refrain from showing them explicitly). Relative errors for the $c_{5,1}$ are of
the order of $0.7\%$ while for the tenth coefficient it is only about
$2\%$. This excellent convergence is not unique for the
$\mathbb{P}_{N,5}^{N+1}(u)$ sequence since for the subdiagonal one,
$\mathbb{P}_{N+1,5}^{N}(u)$, results are basically
indistinguishable. This is so that even a Pad\'e-type sequence
$\mathbb{P}_{0,5}^{N}(u)$ converges nicely and the systematic
prediction of the series coefficients with increasing power $N$ yields
a way to ascribe a theoretical or systematical
error. $\mathbb{P}_{0,5}^{N}(u)$ with $N=2,3$ predict $797.8$
and\ $792.3$ respectively. Taking the difference among them as a way
to estimate the error results into a prediction $792.3 \pm 5.5$ which
nicely agrees with the true coefficient $787.8$. Other results
obtained from these approximants for FOPT and CIPT, as well as the
Borel sum are as impressive as those for the Adler function and we refrain from quoting them
explicitly.

At this point, two comments are in order. If, instead of considering
double poles for the second IR pole and the leading UV pole in
Eq.~(\ref{T5}), we use simple poles the results are worsened.  The
coefficient $c_{5,1}$ obtained from $\mathbb{P}_{1,3}^2(u)$
using $T_3(u) = (u+1)(u-2)(u-3)$, changes from $847.9$ to
$594.1$, to be compared with the exact value 787.8. The reason for this
worsening is simple: fixed poles with wrong exponent forces the
approximant to spend series coefficients to determine the correct
exponents, with an associated  loss of prediction power.
This fact is nicely illustrated by  $\mathbb{P}_{1,3}^2(u)$ in the $\MSb$, which has $(u+1)(u-2)(u-3)(u-3.06)$ as denominator, showing clearly that the Pad\'e must reproduce the double-pole
nature of the second IR renormalon. A scheme variation helps again in this issue since $\mathbb{P}_{1,3}^2(u)$ for $C=-5/3$ has as a denominator
$(u+1.04)(u+1)(u-2)(u-3)$, with the extra pole very close to the first (and dominant) UV pole, emulating its double pole nature.

This shows that simply fixing poles at the correct location but
with the wrong multiplicity does not represent necessarily an
improvement over the situation where the poles are left free. An
intermediate case where only the pole at $u=3$ has an incorrect
exponent does yield improved results compared to the ordinary Pad\'e
approximants. This is in agreement with the findings of
Ref.~\cite{BBJ13} where it was shown that  imposing the correct
structure of the first two leading poles is sufficient to achieve a
good description of the large-$\beta_0$ Adler function. Actually,
in the language of Pad\'e approximants employed here, both the
``reference model'' and  the ``alternative model'' of Ref.~\cite{BBJ13}  can be thought of as a $\mathbb{P}_{0,5}^6(u)$,
i.e., with full denominator fixed in advanced.

The main observation that can be drawn here is that having information
about the first two or three renormalon poles is largely sufficient to
achieve an excellent reconstruction of the series even with only four
coefficients available (this conclusion is in agreement with
Ref.~\cite{BBJ13}). Notice that even though we use large denominators,
we still allow the approximants to have free poles. This helps them to
improve on the convergence as free poles accommodate in an effective
way the rest of the renormalon contributions.


\subsubsection{D-log Pad\'e approximants}
\label{DlogSec}

While information on pole positions yields a clear improvement in the
acceleration of convergence, the precise knowledge on pole positions
can be, in realistic situations,  scarce.  Moreover, knowing the multiplicity of poles is also important. Additionally, some functions present branch point
singularities, which may render their approximation by Pad\'es less
effective.  As discussed in the Introduction, in such situations,
other strategies may be pursuit. In particular, the D-log Pad\'e
approximants of Eqs.~(\ref{Dlog}) and (\ref{DlogNM}) and their extension in Eqs.~(\ref{DlogPP}) and~(\ref{DlogPPNM})
can result optimal.

The original philosophy of the simplest D-log Pad\'e approximant is to
perform a PA not to $f(z)$ but to $F(z) = \frac{\rm d}{{ \rm d}z}
\log[f(z)] $.  Assuming the original function has a singularity at
$z=\mu$ (a pole or a branch cut) the evaluation of the outcome at
$z=\mu$ provides a way to extract, from the series coefficients, the
multiplicity $\gamma$ of the singularity at $\mu$. Here, if $\gamma$
is an integer or not becomes irrelevant which makes the D-log Pad\'es
particularly interesting to approximate functions with branch cuts. Of
course, by unfolding the procedure one obtains the non-rational
approximant $\Dlog_N^M(z)$ of Eq.~(\ref{DlogNM}) to the function
$f(z)$ which, after reexpansion, returns the series coefficients.

\begin{figure}[!t]
\begin{center}\subfigure[Adler function. ]{\includegraphics[width=.52\columnwidth,angle=0]{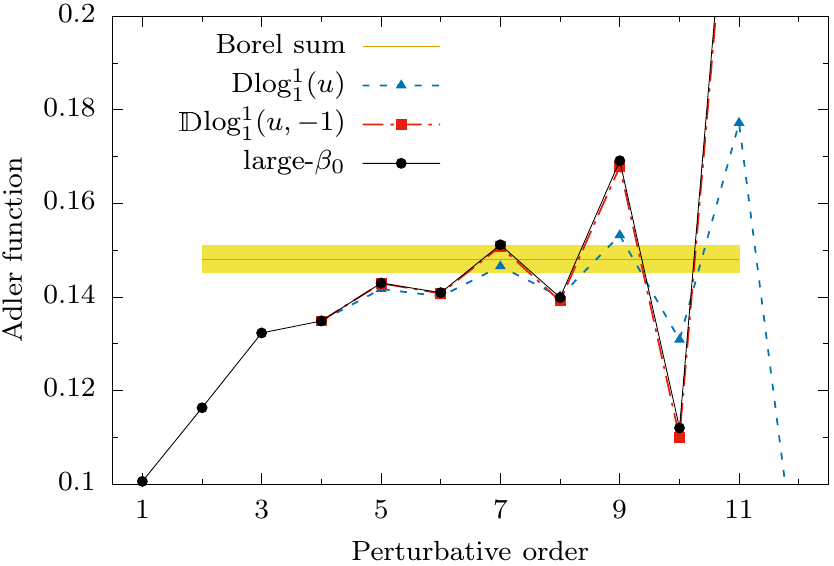}\label{Adler_DLoGP02}}
\subfigure[$\delta^{(0)}$, FOPT.] {\includegraphics[width=.48\columnwidth,angle=0]{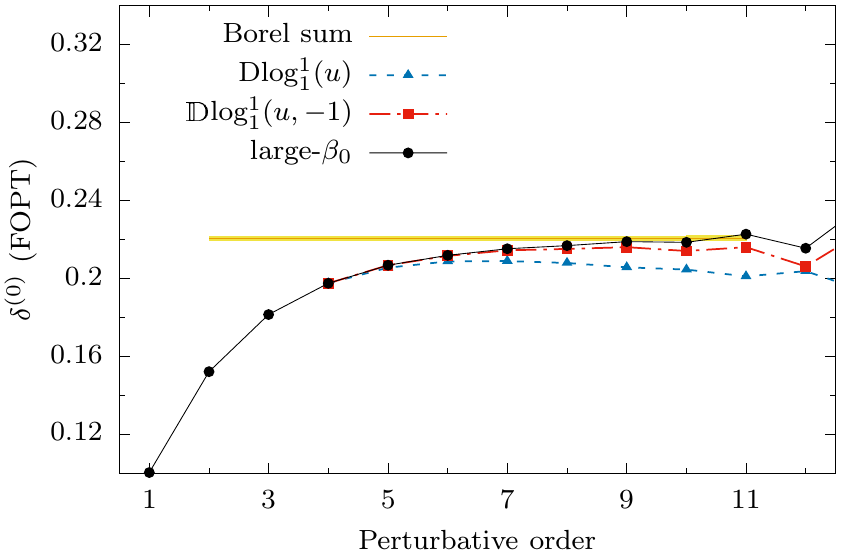}\label{FOPT_DLog_Adler_Lb}}
\subfigure[$\delta^{(0)}$, CIPT.] {\includegraphics[width=.48\columnwidth,angle=0]{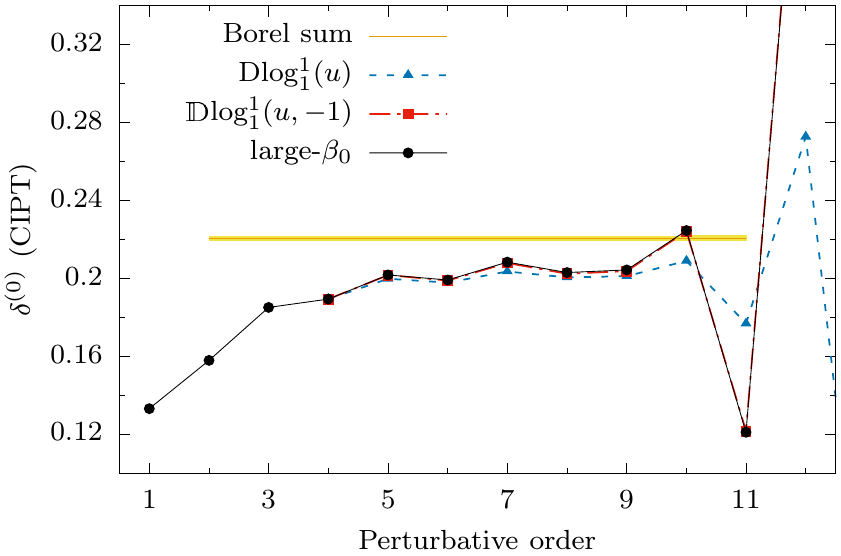}\label{CIPT_DLog_Adler_Lb}}
\caption{Results from the use of the approximants ${\rm Dlog}^1_1(u)$ and  $\mathbb{D}{\rm log}^1_1(u;-1)$ in large-$\beta_0$. (a) Perturbative expansion of the Adler function. (b) $\delta^{(0)}$ in FOPT, and (c)  $\delta^{(0)}$ in CIPT.}
\label{Adler_Dlog}
\end{center}
\end{figure}

The simplest D-log approximant  for the Borel transform of the Adler function, ${\rm Dlog}^{1}_0(u)$,   requires $c_{2,1}$ and $c_{3,1}$ and reads:
\begin{equation}\label{DLexp}
\Dlog^1_0(u) = f(0) e^{ 0.69 u + 1.31 u^2}\, .
\end{equation}
After reexpanding, $\Dlog^1_0(u)$ predicts both $c_{4,1}$ and higher
coefficients and it does it rather accurately, taking into account the
little information that was used. Still, no sign-alternation is
observed.  Going up to $\Dlog^2_0(u)$, the prediction improves a lot
and not only the sign-alternation is now reproduced, but also the
relative error for the $c_{5,1}$ is around $40\%$ (better for the next
$c_{6,1}$ which is just $10\%$ off). At the same order, a $\Dlog^1_1(u)$ yields even
better results, being only $16\%$ off for $c_{5,1}$, with the correct
sign-alternation and a good prediction for $c_{6,1}$ as well, as can be seen in
Fig.~\ref{Adler_Dlog}.  Using
this simplest scenario, with minimum information, the $\Dlog^1_1(u)$
is the approximant that better predicts $c_{5,1}$ among  the
approximants presented in this work so far.

Information on known singularities can be straightforwardly used with
the partial D-log Pad\'es, $\PDlog^M_N(z;\mu)$, of
Eq.~(\ref{DlogPP}).
The singularity closest to the origin in the Borel transform Adler
function is the UV pole at $u=-1$.
We then consider a PA to
$(1+u) \frac{\rm d}{ {\rm d}u} \log[f(u)] $ instead, and construct the
respective $\PDlog$ approximant to $f(u)$. Even with the simplest $\PDlog^1_0(u;-1)$ the sign-alternation of the series coefficient
is recovered, a clear indication of an improvement on the series'
reconstruction. The unfolded approximant reads
\begin{equation}
\PDlog^1_0(u;-1) = f(0) \frac{e^{3.32\, u}}{(1 + u)^{2.6}}
\end{equation}
\noindent
which has a branch point at $u=-1$ but with a good prediction on the actual multiplicity of the first UV renormalon (which is a double pole in large-$\beta_0$).

For the next order, two choices can be considered, the diagonal
$\PDlog^1_1(u;-1)$ and the $\PDlog^2_0(u;-1)$. As it is clear from the
definition in Eq.~(\ref{DlogPP}), $F(z)$ is to a good approximation, a
rational function. Then, the diagonal $\PDlog^1_1(u;-1)$ is the optimal
choice, specially if one is heading towards determining the
multiplicity $\gamma$. In this case, the unfolded approximant reads
\begin{equation}\label{D11}
\PDlog^1_1(u;-1) =\frac{881379.40}{(4.02 -  u)^{10.67} (1+ u)^{1.96}}\, ,
\end{equation}
\noindent
and, upon expansion, predicts the series coefficients for the Borel
transform with unprecedented precision for coefficients up to
$c_{10,1}$, with relative errors amounting to mere $0.8 \%, 6\%, 2\%,
3\%, 2\%, 1\%, 1\%$ for the coefficients $c_{4,1}$ to $c_{10,1}$ in
that order. The excellent results for this approximant can also be seen in Fig.~\ref{Adler_Dlog}. A key point in this extraordinary success is the excellent
reproduction with very little information of the multiplicity of the
first UV pole, as can be seen by the fact that $\gamma = 1.96$ for the
$\mu =-1$ in Eq.~(\ref{D11}), while the other branch cut effectively
emulates the tower of IR poles in the Borel transform.
The result is so good that including  $c_{4,1}$ to construct the $\PDlog^1_2(u;-1)$ yields only a marginal improvement  (predicting  $c_{5,1}$ with
a $5\%$ of relative error, for example).

It is then clear the D-log Pad\'es are able to go much beyond ordinary
PAs in reproducing the main analytical features of the Borel
transformed Adler function in large-$\beta_0$. The success of this approximants
can be understood by examining the function $\frac{\rm d}{{ \rm d}z} \log[f(z)]$.
Let us write the explicit result for the first leading poles, obtained from considering only the first term in the sum of Eq.~(22), 
\beq
F(u) =  \frac{{\rm d}}{{\rm d}u}\log \left(B[\widehat D](u)   \right) = C +\frac{5}{3}- \frac{2}{1+u}+\frac{2}{2-u} +\frac{2}{3-u}+ \cdots
\eeq
This shows that in the D-log Pad\'es the function being approximated is strictly a rational function, with simple poles. The exponential function present in the Borel transform disappears and the approximants do not have to reproduce it. In a sense, the D-logs strategy realizes a situation similar to that of a scheme with $C=-5/3$ which cancels the exponential of the Borel transform but with the additional benefit that $F(z)$ has only simple poles. Their success is, therefore,  not so surprising. 
We can also expect that sequences of PAs that depart significantly from the diagonal will have a slower convergence to $F(u)$ and will, accordingly, lead to a worse approximation of $B[\widehat D](u)$.

Finally, the inclusion of information about the position of more than
one pole can be done in the context of D-log Pad\'es by using a
partial Pad\'e approximant to $F(z)$, with as many poles fixed as one
desires. Using in the denominator of such a PPA $T_7(u) =(2 - u) (3 -
u)^2 (1 + u)^2 (3 + u)^2$, and with $c_{2,1}, c_{3,1}$ and $c_{4,1}$
as input one predicts the series's coefficients with precision better
than $0.1\%$.

In conclusion, the D-log Pad\'es are shown to be an effective way to
improve the convergence of the approximation and gain information
about higher orders. In some sense, they are better than the scheme
transformations since they do not rely on a specific value of $C$.
We turn now to the use of the techniques described here for the function
$\delta^{(0)}_{\rm FO}$ in large-$\beta_0$, before applying them to full QCD.

\subsection[Pad\'e approximants to  $\delta^{(0)}$]{Pad\'e approximants to \boldmath $\delta^{(0)}$}
\label{PAsToDelta0}

We learn from the application of Pad\'e approximants to the Borel
transform of the Adler function that one of the difficulties is the
disentanglement of the leading renormalons. The success of the use of
scheme variations lies partially in this fact, since the method allows
for an enhancement of the leading UV pole with respect to the leading
IR pole at $u=2$. The Borel transform of $\delta^{(0)}$,
Eq.~(\ref{BorelDelta0}), on the other hand, does not have the pole at
$u=2$. The leading UV renormalon is therefore more isolated from the
IR ones. It can be expected that the use of Pad\'e approximants
directly to this Borel transform should yield better and more stable
results than in the case of the Adler function. In this section we
exploit this route. Since  general properties of the Pad\'e
approximants were discussed in the previous section, we will focus here
on the practical problem of forecasting the unknown coefficients given
that only the first four are known exactly --- in order to simulate the case of real QCD.

Let us note that a rational approximant to $\delta^{(0)}$ contains
enough information to allow for a full reconstruction of the Adler function.
The coefficients $c_{n,1}$ can easily be read off from the  FOPT
expansion of $\delta^{(0)}$ as
\begin{align}
  {\delta}^{(0)}_{\text{FO},L\beta} (a_Q) &= \  c_{1,1} \ a_Q + (3.563 \ c_{1,1} + c_{2,1}) \ a_Q^2 + (1.978 \ c_{1,1} + 7.125 \ c_{2,1}+c_{3,1}) \ a_Q^3  \nn \\
  & + (-45.31 \ c_{1,1} + 5.934 \ c_{2,1} + 10.69 \ c_{3,1} + c_{4,1}) \ a_Q^4+\cdots
\end{align}
Additionally, in large-$\beta_0$, Eq.~(\ref{BorelDelta0}) can be used to
extract the Borel transformed Adler function from that of $\delta^{(0)}$.

We start here by applying Pad\'e approximants directly to the series
in $\alpha_s/\pi$, given by Eq.~(\ref{FOPTLb}). 
As we have observed, the  FOPT series in large-$\beta_0$ is rather
well behaved and, at intermediate orders, its asymptotic nature is not
visible yet. This is mapped into a simpler analytic structure in the Borel plane.
 It is therefore likely that in this case the approximation of the series by
Pad\'e approximants in $a_Q$ will lead to a better description than in the case of the Adler function.
In Tab.~\ref{PADelta0},
we display the Adler function coefficients that are obtained from the
application of  Pad\'e approximants directly to the FOPT expansion
of $\delta^{(0)}$. To simulate the situation of real QCD, in the first
two rows, we attempt to forecast $c_{4,1}$, while in the last three we
use $c_{4,1}$ as input. The main observation is that the results for
the coefficients that are not used as input are quite good and
stable. The sign alternation of the series is correctly predicted by
all the Pad\'es and good consistency between the results using three
and four coefficients as input is achieved. The results are therefore
quite remarkable.

Let us examine one of the PAs in more detail. For $P_1^2$, which uses only the first three coefficients as input, we find
\beq
P_1^2(a_Q) =  \frac{0.1779\,a_Q-0.0895\,a_Q^2}{0.1779-a_Q}.
\label{P12Delta0}
\eeq
First, we see that the series obtained from such a Pad\'e is convergent (since $a_\tau\approx 0.1$) and
does not reproduce the factorial growth of the coefficients.
The pole that appears around $a_Q\approx 0.18$ is, however, rather stable --- it does not seem to be
transient in nature and is present in all the approximants to  ${\delta}^{(0)}_{\text{FO},L\beta}$. It may be worth noting that the pole is in the IR
and corresponds to a scale of $\sim 650$ MeV. It may, therefore,  be related
to IR physics but we refrain from further speculation regarding the nature of this pole. An estimate for the
sum of the series can be obtained simply using $a_\tau = 0.316/\pi$ in Eq.~(\ref{P12Delta0}) to give $P_1^2(a_\tau)=0.2198$. This result is also in very good agreement with the value obtained from the Borel integral of the exact large-$\beta_0$ result, which gives $0.2208\pm 0.0039$. Even better results can be obtained
from the PAs that use four coefficients as input, as shown in the last three rows of Tab.~\ref{PADelta0}.

\begin{table}[!t]
  \begin{center}{
  \caption{Adler function coefficients extracted from PAs $P_M^N(a_Q)$ to the FOPT $\alpha_s/\pi$ expansion  of $\delta^{(0)}$ in the large-$\beta_0$ limit. }
\begin{tabular}{ccccccccc}
  \toprule
  & $c_{4,1}$ & $c_{5,1}$ & $c_{6,1}$ & $c_{7,1}$ &   $c_{8,1}$ & $c_{9,1}$  \\
  \midrule
  \rowcolor{light-gray} Large-$\beta_0$ (exact) & $24.83$ &   $787.8 $      & $-1991$    &   $9.857\times 10^{4}$  &  $-1.078\times10^6$  &  $2.775\times10^7$   \\
  $P^2_1(a_Q)$  &   $29.95$     &   $723.7$    &   $-703.4$    &   $7.405\times 10^4$   & $-5.871\times 10^{5}$  & $1.649\times 10^{7}$    \\
  $P^1_2(a_Q)$  &   $28.66$     &   $728.2$    &   $-874.3$    &   $7.554\times 10^4$   & $-6.224\times 10^{5}$  & $1.703\times 10^{7}$    \\
  $P^3_1(a_Q)$  &    input      &   $740.0$    &   $-1363$     &   $7.956\times 10^4$   & $-7.211\times 10^{5}$  & $1.851\times 10^{7}$    \\
  $P^2_2(a_Q)$  &    input      &   $749.3$    &   $-1444$     &   $8.169\times 10^4$   & $-7.514\times 10^{5}$  & $1.917\times 10^{7}$    \\
  $P^1_3(a_Q)$  &    input      &   $743.6$    &   $-1393$     &   $8.035\times 10^4$   & $-7.321\times 10^{5}$  & $1.875\times 10^{7}$    \\
  \bottomrule
\end{tabular}
\label{PADelta0}
}\end{center}
\end{table}

We now turn to PAs to the Borel transformed $\delta^{(0)}$. The
results are improved when compared with PA to $B[\widehat D]$,
although the improvement is far from spectacular. For $c_{5,1}$ we
find the values $263.9$, $603.6$, and $1024$ from $P_2^1$, $P_3^0$,
and $P_1^2$ respectively, to be compared with $52.33$, $560.9$, and
$1770.1$ from the same PAs to $B[\widehat D]$. Clearly, one must
resort to the other methods discussed in the previous section in order
to accelerate the convergence. Again, D-Log Pad\'es turn out to be the
optimal way to improve the convergence while remaining completely
model independent. Their success can be understood from the
a study of the function $F(u)=\frac{{\rm d}}{{\rm d}u}\log \left(B[\delta^{(0)}](u)   \right)$, as is in the case of the Adler function.
Here we find, retaining only the first term in the sum of Eq.~(\ref{BDLb}),
\begin{align}
  F(u) & =  \frac{{\rm d}}{{\rm d}u}\log \left(B[\delta^{(0)}](u)   \right) \nn \\
  &= C +\frac{5}{3}+\pi\cot(\pi u) -\frac{2}{1+u} +\frac{3}{3-u} +\frac{1}{4-u}+\frac{1}{1-u}+\frac{1}{2-u}-\frac{1}{u}+ \cdots
\end{align}
The leading analytic structure of $F(u)$  is now even simpler than for the Adler function. The poles at $u=0$, $u=1$, and $u=2$ are exactly cancelled by the presence of $\pi\cot(\pi u)$ leaving only a leading UV pole at $u=-1$, an IR pole at $u=3$ and a subleading IR pole at $u=4$.\footnote{To get a non-vanishing residue for the pole at $u=4$ one must add the second term in the sum of Eq.~(\ref{BDLb}).} It is therefore expected that the D-log Pad\'es should perform even better in the present case.

We present results for the D-Log Pad\'es applied to
$B[\delta^{(0)}]$  in Tab.~\ref{DLogDelta0}.  These results
also represent an improvement when compared to those of
Sec.~\ref{DlogSec}. The predictions for $c_{5,1}$ have a much smaller
relative error, a factor of 2 to 5 times smaller than those obtained
from the PAs to the Borel transformed Adler function. The sign
alternation is well reproduced by the Pad\'es with four coefficients
used as input and their Borel integral provide excellent estimates for
the true value of the series (we find, e.g., 0.2199 from
$\text{DLog}^1_1(u)$). However, one must note that the results from the
D-Log Pad\'es applied to $B[\delta^{(0)}]$ are less good than those of
Tab.~\ref{PADelta0}. For example, the coefficient $c_{4,1}$ is wrong
by a factor of about two while in Tab.~\ref{PADelta0} it is only a few
percent off. Nevertheless, the description of the Borel transformed
$\delta^{(0)}$ by D-Log Pad\'es has the advantage that the factorial
growth of the coefficients is reproduced and an asymptotic series is obtained,
in line with the exact result.

We have checked that the results discussed in this section can be
further improved by using information about the renormalons. For
example, imposing the existence of the leading UV pole at $u=-1$
through the use of $\PDlog_1^1(u;-1)$ leads to an almost exact
reproduction of the series. We prefer, however, to remain as model
independent as possible and we chose to focus here on the results
obtained from the most model independent methods (PAs and D-log
Pad\'es). By using $\delta^{(0)}$ and its Borel transformed, these
model-independent methods lead to results as good as those obtained
from the Adler function imposing information on the renormalons.

\begin{table}[!t]
  \begin{center}{
  \caption{Adler function coefficients extracted from D-Log Pad\'e approximants, ${\rm Dlog}_M^N(u)$, constructed to  $B[\delta^{(0)}](u)$ in the large-$\beta_0$ limit. }
	\begin{tabular}{ccccccccc}
		\toprule
			& $c_{4,1}$ & $c_{5,1}$ & $c_{6,1}$ & $c_{7,1}$ &   $c_{8,1}$ & $c_{9,1}$  \\
		\midrule 
	\rowcolor{light-gray} 	Large-$\beta_0$ (exact) & $24.83$ &   $787.8 $      & $-1991$    &   $9.857\times 10^{4}$  &  $-1.078\times10^6$  &  $2.775\times10^7$   \\
			$\text{Dlog}^1_0(u)$  &   $41.84$     &   $756.1$    &   $848.6$    &   $7.453\times 10^4$    & $-3.284\times 10^{5}$  & $1.498\times 10^{7}$    \\
		$\text{Dlog}^0_1(u)$  &   $44.43$     &   $776.1$    &   $1294$     &   $7.778\times 10^4$    & $-2.502\times 10^{5}$  & $1.538\times 10^{7}$    \\
		$\text{Dlog}^2_0(u)$  &    input      &   $650.2$    &   $-1824$    &   $6.319\times 10^4$    & $-7.470\times 10^{5}$  & $1.545\times 10^{7}$    \\
		$\text{Dlog}^1_1(u)$  &    input      &   $818.7$    &   $-2738$    &   $1.189\times 10^5$    & $-1.663\times 10^{6}$  & $4.495\times 10^{7}$    \\
		$\text{Dlog}^0_2(u)$  &    input      &   $594.9$    &   $-1974$    &   $5.560\times 10^4$    & $-6.796\times 10^{5}$  & $1.432\times 10^{7}$    \\
		\bottomrule
	\end{tabular}
        \label{DLogDelta0}
}\end{center}
\end{table}

\begin{figure}[!t]
\begin{center}
\subfigure[$\delta(0)$ from $P_2^2(a_Q)$.] {\includegraphics[width=.48\columnwidth,angle=0]{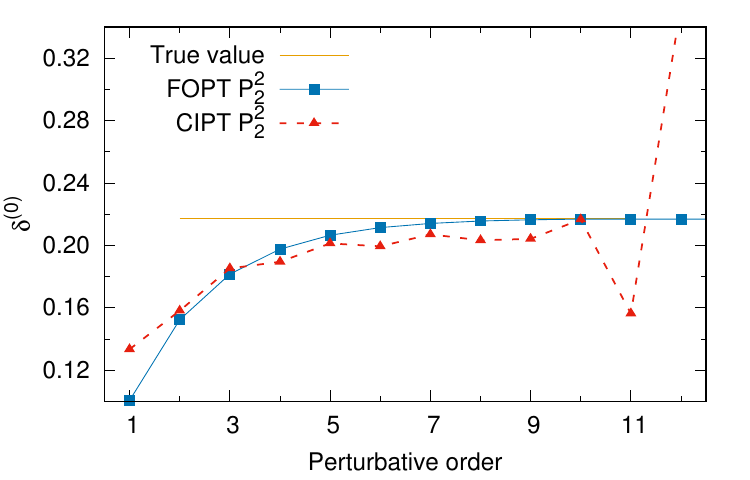}\label{Delta0P22}}
\subfigure[$\delta(0)$ from ${\rm Dlog}_1^1(u)$.] {\includegraphics[width=.48\columnwidth,angle=0]{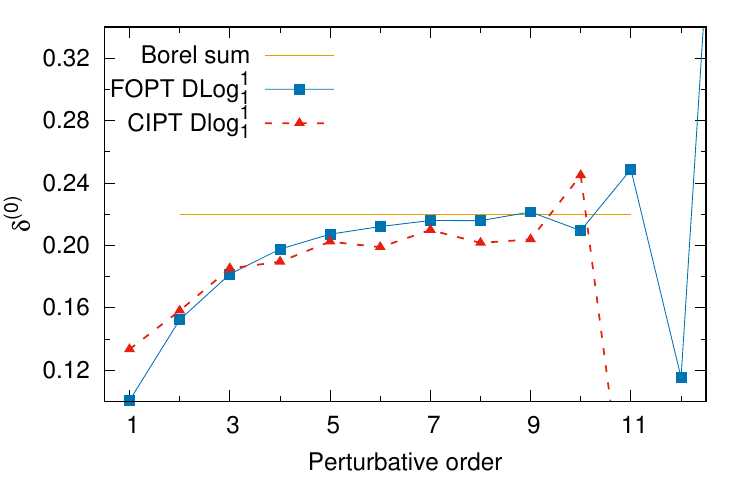}\label{Delta0Dlog11}}
\caption{Perturbative expansion of $\delta^{(0)}$ in FOPT and CIPT obtained from (a) $P_2^2(a_Q)$ (5th row of
Tab.~\ref{PADelta0}) and (b) $\Dlog_1^1(u)$ (5th row of
Tab.~\ref{DLogDelta0}).}
\label{Delta0PadesLb}
\end{center}
\end{figure}

We close this section with a visual account of the results discussed
here. In Fig.~\ref{Delta0P22}, we display the $\delta^{(0)}$ FOPT and
CIPT expansions obtained from the approximant $P_2^2(a_Q)$, for which
we show the coefficients up to $c_{9,1}$ in the 5th row of
Tab.~\ref{PADelta0}. The main observation is that, up to the 9th order,
the result is strikingly similar to the exact ones (see, e.g., the
black lines in Figs.~\ref{FOPT_DLog_Adler_Lb} and
\ref{CIPT_DLog_Adler_Lb}). However, the FOPT series thus obtained is convergent, and
after the 8th order its result stabilizes around the sum of the series, which
cannot happen in the exact results. This does not prevent the
description from being excellent at intermediate orders, since the
true value of the convergent series is very similar to the central
value of the Borel integral of the exact results. For CIPT, similar
observations hold and up the 9th order the result is almost
indistinguishable from the exact ones. In particular, the fact that
FOPT is the best approximation in large-$\beta_0$ is well
reproduced. In Fig.~\ref{Delta0Dlog11}, we display results for
$\Dlog_1^1(u)$, for which the coefficients are given in the 5th row of
Tab.~\ref{DLogDelta0}. Again, visually, the results are almost identical
to the exact ones. Now, both series are divergent, as is the case in
the exact results, and this feature shows up after the 9th order,
although the divergence, here, is more pronounced than in the exact
large-$\beta_0$ case. In both cases, however, the approximants provide
a good estimate for the first few unknown coefficients and give an
excellent account of the series even though no information about
the renormalons was used, in contrast with some of the results of
Fig.~\ref{FOPT_DLog_Adler_Lb} and~\ref{CIPT_DLog_Adler_Lb}).


\subsection{Summary and discussion}

Here we summarize the main conclusions that can be drawn from the
application of Pad\'e theory to the results in the large-$\beta_0$
limit of QCD.

The application of the Pad\'e approximants to the $\MSb$ Borel
transformed Adler function in the large-$\beta_0$ limit displays
convergence. With six or seven coefficients one is able to reproduce
very well all the essential aspects of the original series: its higher
order coefficients, its Borel sum, and even the position of the
dominant renormalon poles. There are, however, outliers and it can happen that
the next Pad\'e in a given sequence does not make the results
better. The existence of these outliers is well understood in the
theory of Pad\'es and we have been able to show for specific examples
why this happens. All in all, convergence is always observed once an
even larger number of parameters is added.  Results using only four
coefficients of the series to construct the Pad\'es --- which
corresponds to the number of available coefficients in QCD --- are,
however, far from spectacular.  In the absence of more coefficients,
one must exploit strategies to improve the approximation of the
original function.

  In the case of the large-$\beta_0$, we have shown that using less
  perturbative schemes, which with our conventions corresponds to
  $C<0$, one is able to construct better approximants with the same
  number of parameters. This improvement is due to the fact that the
  Borel transform in these schemes becomes largely dominated by the
  leading UV pole: one observes the sign alternation earlier and the
  Pad\'es can easily reproduce the main features of the analytical
  structure of the series. The specific choice $C=-5/3$ has the
  additional advantage of removing the term $e^{C+5/3}$ from the Borel
  transform, which makes it a strict rational function. As a
  by-product, we are able to devise a strategy to reveal the effects
  of the dominant UV pole. Constructing the series for $C<0$ the
  sign-alternation of the coefficients sets in much earlier, already
  from the second coefficient. 

  Next, we have seen that the use of partial Pad\'e approximants,
  where the available information about the renormalon poles can be
  exploited, leads to significant improvements. However, it is
  important to use this information with care, since imposing the
  existence of a pole but with wrong multiplicity forces the Pad\'e to
  ``spend'' series coefficients fixing the analytic structure of the
  given pole. If the information of the leading UV pole is used
  correctly, this leads to a significant improvement on the
  results. Finally, we find that imposing the structure of the first
  two leading renormalons leads to an almost perfect reproduction of
  the Adler function. This is in line with the renormalon models of
  Ref.~\cite{BBJ13}.

  We then investigated the use of D-log Pad\'es. These are an
  interesting alternative having in mind applications to QCD, since
  they do not require that the function be meromorphic. Their
  application in large-$\beta_0$ was very successful and we can safely
  conclude that, given the limited information available, they are
  the best way to accelerate convergence of the procedure. We
  highlight the use of partial D-log Pad\'es where only the existence
  of first UV pole is used as input. The results of such partial
  D-logs are truly impressive and lead to an almost exact
  reproduction of the Adler function in large-$\beta_0$ with an almost
  model-independent method. We were able to explain their success in
  terms of the analytic structure of the Borel transformed Adler function.

  We then turned to approximants constructed to $\delta^{(0)}$. First,
  we have shown that PAs to the FOPT series in $\alpha_s/\pi$,
  contrary to the case of the Adler function, do lead to a very good
  reconstruction of the series at intermediate orders.  In the case of
  $\delta^{(0)}$, since the FOPT series is rather well behaved and
  regular up to the 10th order (with terms that are consistently
  smaller than their predecessor) the approximation
  obtained from these Pad\'es is very reasonable. We have then shown
  that good results can also be obtained from D-log Pad\'e
  approximants constructed to $B[\delta^{(0)}]$.  They are completely
  model independent and have the advantage that the factorial growth
  of the coefficients is automatically implemented.  Both methods lead
  to good predictions for the true value of the series and are
  sufficient to conclude that FOPT is the best prescription in
  large-$\beta_0$. Again, the success of the method can be explained
  by the simpler analytic structure of the Borel transformed
  $\delta^{(0)}$ since it does not have the pole at $u=2$ and all
  other poles become simple (with the exception of the ones at $u=3$
  and $u=4$).

  The systematic study performed in the large-$\beta_0$ limit leads to
  strategies for impressive determinations of the higher order
  coefficients and the Borel sum of the Adler function and
  $\delta^{(0)}$ series. We were able to find methods to unravel
  dominant and subdominant poles, as well as to reorganize the
  available information in order to optimize the approximation by PAs
  and its variants. With these strategies at hand we can now perform
  a similar analysis in full QCD and present our final results.


\section{The QCD case}
\label{sec:QCD}

In this section we will apply the techniques developed and tested 
in large-$\beta_0$ to the real case of QCD. Let us first remind what
is known in this case. In perturbation theory, the Adler function
expansion is known to five loops, hence to order
$\alpha_s^4$~\cite{BCK08}. We would like to obtain predictions for the
coefficients $c_{5,1}$ and higher. The first four coefficients of the Adler function in QCD,
displayed in Eq.~(\ref{DinQCD}), already show a significant deviation
from the large-$\beta_0$ results, although the coefficients of the
$\delta^{(0)}$ FOPT expansion, Eq.~(\ref{FOPTQCD}), are closer to the their large-$\beta_0$ counterparts. 
The differences between full QCD and the large-$\beta_0$ limit also
show up in the Borel transform.  In QCD, we know that the Borel
transform is no longer a meromorfic function. Although the renormalon
singularities remain at the same position, anomalous dimensions of the
operators and higher-order $\beta$-function coefficients now change their nature from poles to branch points. What is
more, at  every branch point there are confluent
singularities~\cite{Renormalons,BJ08}. Experience shows that Pad\'e approximants can be safely
employed to approximate functions with branch cuts, but we no longer have
convergence theorems to exploit, without further knowledge on the properties of
these branch cuts.

Let us start with the simplest approximants: PAs constructed directly
to the $\alpha_s/\pi$ expansion of the Adler function.  We have
discussed that in large-$\beta_0$ this was not the optimal strategy. Given that we have now only the first four coefficients, we
start by building the PAs that would ``postdict'' the last known term of
the series, $c_{4,1}$. From the approximants $P_1^2(a_Q)$ and
$P_2^1(a_Q)$ we obtain the five-loop coefficients with $51\%$ and
$67\%$ relative error, respectively. The coefficient $c_{5,1}$ is
predicted to have quite low values $96.2$ and $50.5$. Our experience
from large-$\beta_0$ already signals that this strategy is not
optimal.  If we construct approximants that now use $c_{4,1}$ as input
we extract $c_{5,1}$ coefficients that can differ from the previously
obtained by a factor of 9, indicating that the procedure is very
unstable.

We then turn to the approximation of the Borel transformed Adler
function, which proved to be more stable in large-$\beta_0$. Again, we
first try to obtain a ``postdiction'' of $c_{4,1}$, using 
$P_1^1(u)$ and $P_2^0(u)$.  The value of $c_{4,1}$ thus obtained has a
relative error of about $26\%$, which is an improvement with respect
to the previous method, and leads to predictions of $c_{5,1}\sim
280$. However, when we construct approximants that include the true
value of $c_{4,1}$ as input, $P_1^2$, $P_2^1$, $P_3^0$, there is no
sign of stability. The predictions for $c_{5,1}$ and higher order
coefficients change substantially which, from our experience in
large-$\beta_0$, signals that the approximants are not optimal.  That
only four coefficients is not enough for the usual Pad\'e approximants
to perform a stable prediction of the higher orders comes as no
surprise, since this is what happens in large-$\beta_0$ even though
the analytical structure of the Borel transform is simpler there. We
must again resort to methods for the  acceleration of the convergence of the
procedure.

We start by investigating other renormalization schemes. Scheme
variations in full QCD can also be performed using a single parameter $C$, through the generalization
of Ref.~\cite{BJM16} (to which we refer for further
details). Based on the lessons from large-$\beta_0$ we rewrite the
Adler function series in schemes with negative $C$, hence with larger
values of the coupling. As an example, let us take again the value
$C=-5/3$. The first four coefficients of the Adler function are now
\beq
\widehat D^{(C=-5/3)} (a_Q) = \hat a_Q  -2.110 \,\hat a_Q^2 + 2.779 \,\hat a_Q^3 + 19.87 \,\hat a_Q^4 +\cdots
\eeq
There are marked differences in the coefficients when compared with
Eq.~(\ref{DCm2}). In contrast to the large-$\beta_0$ case, the series
in QCD with negative $C=-5/3$ does not display a systematic sign
alternation. It seems, therefore, that the UV singularity is less dominant
than in large-$\beta_0$ and the pattern of signs indicate a more
complicated interplay between the poles. Possibly, the IR poles give a
larger contribution to the QCD series relatively to the
large-$\beta_0$ case. This is in line with the results of the models
of Refs.~\cite{BJ08,BBJ13}. In these models, the IR poles give larger
contributions at intermediate orders and the systematic sign
alternation sets in only at the 11th order. Since we are not able to
unequivocally suppress the contribution of the IR poles, the strategy
of using scheme variations to optimize the use of Pad\'e approximants
does not turn out optimal in QCD (although it corroborates, in part,
the results of Refs.~\cite{BJ08,BBJ13}).

We resort then to the use of D-log Pad\'es to obtain approximants to $B[\widehat D](u)$. Their use in full QCD is also well motivated, since they are
designed for functions that have branch cuts. The use of D-log Pad\'es
to postdict $c_{4,1}$ leads to values with relative error of $54\%$
[$\Dlog^1_0(u)$] and $21\%$ [$\Dlog_1^0(u)$] which again signals that
the procedure is not optimal. The use of an additional coefficient as input
does not change this picture significantly, since it leads to unstable
results for higher-order coefficients, which is understandable again due to the presence of confluent singularities.

Based on our experience from the large-$\beta_0$ limit, one is the led
to conclude that model-independent approximants constructed to the
Borel transformed Adler function are not robust enough in QCD with
only the first four coefficients available. In large-$\beta_0$ the
knowledge about renormalon singularities could be used to optimize the
predictions. Here, however, we face additional difficulties. First,
the available knowledge is more scarce. Only the first few renormalons
(the leading UV, and the two leading IR) have had their branch cut
structure investigated in detail~\cite{BJ08,BHJ15}. Second, the fact
that now the renormalons become confluent singularities renders
much more difficult the use of the available knowledge to devise
optimized approximants. And, finally, it would be desirable to remain
as model independent as possible.  We therefore turn directly to the
most successful model-independent strategy devised in large-$\beta_0$:
the use of the FOPT series for $\delta^{(0)}$.

In large-$\beta_0$  approximants constructed to
$\delta^{(0)}_{\rm FO}$ and $B[\delta^{(0)}] $ resulted optimal. The
perturbative series for $\delta^{(0)}_{\rm FO}$ in large-$\beta_0$ and
in QCD have rather similar coefficients. This means that the
regularity of the series is preserved in QCD, which suggests that it
can be well approximated by Pad\'e approximants constructed directly
to the series in $\alpha_s/\pi$. Furthermore, although
Eq.~(\ref{BorelDelta0}) is strictly valid only in large-$\beta_0$,
because it relies on the one-loop running of the coupling,
modifications to this result would be solely due to higher-orders in the
$\alpha_s$ evolution. We can therefore expect that a suppression of
the leading IR singularity at $u=2$, as well as a suppression of all
the other renormalons except for the ones at $u=3$ and $u=4$, would
survive in full QCD and render this Borel transform more amenable
to approximation by rational functions.

We start with Pad\'e approximants applied to the $\alpha_s/\pi$
expansion of $\delta^{(0)}_{\rm FO}$. As before, we begin with a
post-diction of $c_{4,1}$ using $P_2^1(a_Q)$ and $P^2_1(a_Q)$. The
results for six higher-order coefficients obtained from these
approximants are shown in Tab.~\ref{Delta0PadesQCD}. The relative
error from the central values of $c_{4,1}$ is now $\sim 13\%$. This is
quite remarkable when put into perspective since before the true value
of $c_{4,1}$ was computed, a forecast of this coefficient using other
methods and including additional information (taking into account
known terms of order $\alpha_s^4 N_f^3$ and $\alpha_s^4N_f^2$) yielded
$c_{4,1}=27\pm 16$~\cite{BCK02,KS95,Kataev:1995vh}, a central value which was $45\%$
off. This gives an idea of how powerful optimal PAs can be.

\begin{table}[!t]
     \begin{center}{
  \caption{QCD Adler function coefficients  from  PAs constructed to the $\alpha_s$ expansion of  $\delta^{(0)}_{\rm FO}$. }
		\begin{tabular}{ccccccccc}
		\toprule
			& $c_{4,1}$ & $c_{5,1}$ & $c_{6,1}$ & $c_{7,1}$ &   $c_{8,1}$ & $c_{9,1}$ &  Pad\'e sum  \\
		\midrule
		$P^2_1$  &   $55.62$     &   $276.2$    &   $3865$      &   $1.952\times 10^4$   & $4.288\times 10^{5}$   & $1.289\times 10^{6}$  & 0.2080  \\
		$P^1_2$  &   $55.53$     &   $276.5$    &   $3855$      &   $1.959\times 10^4$   & $4.272\times 10^{5}$   & $1.307\times 10^{6}$  & 0.2079  \\
		$P^3_1$  &    input      &   $304.7$    &   $3171$      &   $2.442\times 10^4$   & $3.149\times 10^{5}$   & $2.633\times 10^{6}$  & 0.2053  \\
		$P^1_3$  &    input      &   $301.3$    &   $3189$      &   $2.391\times 10^4$   & $3.193\times 10^{5}$   & $2.521\times 10^{6}$  & 0.2051  \\
		\bottomrule
	\end{tabular}
     \label{Delta0PadesQCD}
     }
\end{center}
\end{table}

Inspecting the Pad\'e approximants of the first two rows of
Tab.~\ref{Delta0PadesQCD} they reveal a pole around $a_Q=0.1973$, of
similar nature to the one found in large-$\beta_0$. Additionally,
$P_2^1$ has a pole far from the origin at $a_Q= 7.25$. This makes
their expansion around $a_Q=0$ similar and their predictions
turn out to be almost degenerate. Therefore, a stronger  test for
stability comes with the use of $c_{4,1}$ as input, to obtain
$c_{5,1}$ and higher. One could construct, for example, the approximant $P_2^2$. However,
this approximant has a defect, in the sense discussed in Sec. \ref{MSbLb}, a pole and a zero that cancel almost exactly, effectively
reducing the order of the approximant. Although its results are not completely inconsistent (e.g., $c_{5,1}$ turns out to be $242$)
we have shown that it is wise to avoid approximants of this type and we will
discard $P_2^2$. 
We turn then to the results obtained for $P_1^3$ and
$P_3^1$  which are also shown in
Tab.~\ref{Delta0PadesQCD}. Now, the forecasts of $c_{5,1}$ are 304.7 and 301.3 respectively. We can note a striking stability of
the results for $c_{5,1}$ and $c_{6,1}$; even $c_{7,1}$ and $c_{8,1}$
are remarkably similar in all of the four approximants considered.
The use of the PAs to sum the asymptotic series also leads to consistent
result in  all cases, as can be
seen in the last column of Tab.~\ref{Delta0PadesQCD}.
Our experience from large-$\beta_0$ indicates that this stability and the good
prediction of $c_{4,1}$ strongly corroborate the robustness of the results.
We have checked that the use of D-log Pad\'e approximants is also very successful. 
We are, therefore, in a position to conclude that using PAs to
$\delta^{(0)}_{\rm FO}$ in QCD is at least as stable as in
large-$\beta_0$. We should then investigate the approximants
constructed to its Borel transformed.

As in the previous section, the quality of the forecast of $c_{4,1}$
as well as stability arguments lead us to conclude that the D-log
Pad\'es are the optimal approximants to
$B[\delta^{(0)}](u)$. Higher-order coefficients obtained from D-log
Pad\'es constructed to $B[\delta^{(0)}](u)$ in QCD are shown in
Tab.~\ref{Delta0BorelPadesQCD}. Now, the postdiction of the last known
coefficient, $c_{4,1}$, has a relative error of only about $\sim 6\%$,
about half of what was obtained with Pad\'es to the series in
$\alpha_s$. Also, the stability of the results when using the exact
value of $c_{4,1}$ as input is quite remarkable. The results for
$c_{5,1}$ and $c_{6,1}$ are rather stable not only among the D-log
Pad\'es of Tab.~\ref{Delta0BorelPadesQCD} but also when compared with
the results of Tab.~\ref{Delta0PadesQCD}. The   approximant,
$\Dlog^1_1$, not shown in Tab.~\ref{Delta0BorelPadesQCD}, leads to slightly lower values for the coefficients (e.g., $c_{5,1}= 237$), but
even these apparent instability can again be understood in terms of a partial cancelation between a pole and a  zero present in the $P_1^1$
used for its construction. We, therefore, consistently discard this approximant. 
 It
is also interesting to observe that all the D-log Pad\'es of
Tab.~\ref{Delta0BorelPadesQCD} predict that the sign alternation of
the series starts at order 11.  This is in
agreement with the speculation we advanced, based on scheme
variations, that the UV singularity in QCD is less prominent which
should postpone the sign alternation with respect to large-$\beta_0$
where it sets in from $c_{6,1}$ on.  Finally, the Borel sum of the
series obtained from these D-log Pad\'es is also very consistent (last
column of Tab.~\ref{Delta0BorelPadesQCD}).

\begin{table}[!t]
   \begin{center}{
  \caption{QCD Adler function coefficients  from D-Log Pad\'e approximants formed  to  $B[\delta^{(0)}](u)$. }
	\begin{tabular}{ccccccccc}
		\toprule
		& $c_{4,1}$ & $c_{5,1}$ & $c_{6,1}$ & $c_{7,1}$ &   $c_{8,1}$ & $c_{9,1}$ & Borel sum\\
		\midrule 
		$\text{DLog}^1_0$  &   $51.90$     &   $272.6$    &   $3530$     &   $1.939\times 10^4$    & $3.816\times 10^{5}$   & $1.439\times 10^{6}$  & 0.2050  \\
		$\text{DLog}^0_1$  &   $52.08$     &   $273.7$    &   $3548$     &   $1.953\times 10^4$    & $3.840\times 10^{5}$   & $1.456\times 10^{6}$  & 0.2052  \\
		$\text{DLog}^2_0$  &    input      &   $254.1$    &   $3243$     &   $1.725\times 10^4$    & $3.447\times 10^{5}$   & $1.187\times 10^{6}$  & 0.2012 \\
		$\text{DLog}^0_2$  &    input      &   $256.4$    &   $3271$     &   $1.769\times 10^4$    & $3.493\times 10^{5}$   & $1.258\times 10^{6}$  & 0.2019 \\
		\bottomrule
	\end{tabular}
     \label{Delta0BorelPadesQCD}
     }
\end{center}
\end{table}

The picture that emerges from the results of this section is that the
use of $\delta^{(0)}_{\rm FO}$ and its Borel transform lead to the best model-independent
approximants in QCD --- as is the case in large-$\beta_0$.  The
quality of the predictions of $c_{4,1}$ as well as the stability of
the results among different approximants signal that we have managed
to obtain a robust description of $\delta^{(0)}$ and of the Adler function at higher
orders. In the next section, we extract our final results and perform
error estimates.


\subsection{Final results and error estimates}

In producing our final results, we will try to remain as conservative as
possible. We extract our final estimates for the higher-order
coefficients from the eight approximants of Tabs.~\ref{Delta0PadesQCD}
and~\ref{Delta0BorelPadesQCD} including, thus, those that have only
three coefficients as input parameters. By doing so, we take advantage
of Pad\'es that belong to different sequences and can obtain a more
reliable error estimate for our final coefficients. Since one of the
most striking features of these results is their stability, we will
not try to favour one approximant over another, even though one could
try to inspect their analytic structure in detail with this goal in mind.  Our
final estimate of the coefficients and of the true value of
$\delta^{(0)}$ is obtained as the average of the eight results of
Tabs. \ref{Delta0PadesQCD} and~\ref{Delta0BorelPadesQCD}.
To these averages we add an error equal to the maximum spread
found between the coefficients obtained from two different approximants.
This error should certainly not be interpreted in a statistical sense; it
gives an interval  where the value of the coefficient is expected to lie.

This procedure applied to the six-loop coefficient, $c_{5,1}$, leads to
\beq
c_{5,1} =  277 \pm 51,\label{finalc51}
\eeq
which largely covers all the results obtained from our optimal
approximants.  Therefore, in a sense, our error estimate could even be
considered as too conservative --- even if much smaller than other
estimates in the literature. For example, in Ref.~\cite{BJ08} the estimate $c_{5,1}
=283 \pm 142$ is used, while in Ref.~\cite{BCK02} one finds $c_{5,1}=
145\pm 100$ (using only partial information about the five-loop coefficient). The value obtained from the principle of Fastest Apparent Convergence (FAC) in Ref.~\cite{BCK08} is $c_{5,1}=275$, remarkably close to our final central value, given in Eq.~(\ref{finalc51}).

On the basis of what we know about the series
  coefficients, it seems extremely unlikely that the six-loop
  coefficient would not be within these bounds.

Results for coefficients $c_{6,1}$ and higher are given in
Tab.~\ref{FinalCoeff}. The final values for the Adler function
coefficients are extracted with reasonable errors up to
$c_{10,1}$. One should remark that due to the $\alpha_s$ suppression
at these higher orders, an error that seems large in the coefficient
does not translate into a very large uncertainty in the sum of the
series. The situation changes only for $c_{11,1}$. For this
coefficient, six of the PAs of Tabs.~\ref{Delta0PadesQCD}
and~\ref{Delta0BorelPadesQCD} predict that the sign alternation  sets in. However, two of the approximants do not, which
leads to the huge error. Therefore, we find some indication that the
sign alternation of the Adler function coefficients sets in at the
eleventh order (in agreement with \cite{BJ08}). This agrees
with our expectation that the UV singularity in QCD should be less dominant than
in large-$\beta_0$. This instability signals that our results cease to be
fully reliable at the  11th order.

\begin{table}[!t]
\begin{center}
  {
     \caption{Final values for the QCD Adler function coefficients  obtained from  PAs to  $\delta^{(0)}_{\rm FO}$. }
	\begin{tabular}{cccc}
		\toprule
			 $c_{5,1}$ & $c_{6,1}$ & $c_{7,1}$ &   $c_{8,1}$   \\
		       $277\pm 51$   &   $3460\pm 690$    &  $(2.02 \pm 0.72)\times10^4$    &   $(3.7\pm 1.1)\times 10^5$   \\
		\midrule 
			 $c_{9,1}$ & $c_{10,1}$ & $c_{11,1}$ &   $c_{12,1}$   \\
		       $(1.6\pm 1.4)\times10^6$   &   $(6.6\pm 3.2)\times10^7$    &  $(-5\pm 57)\times10^7$    &   $(2.1\pm 1.5)\times 10^{10}$   \\
		\bottomrule
	\end{tabular}\label{FinalCoeff}
}
\end{center}
\end{table} 

We apply the same procedure described above to obtain an estimate for
the true value of the $\delta^{(0)}$ using the results in the last
columns of Tabs.~\ref{Delta0PadesQCD} and~\ref{Delta0BorelPadesQCD}.
Using $\alpha_s(m_\tau^2) = 0.316 \pm 0.010$~\cite{PDG}, this leads to
\beq
\delta^{(0)} = 0.2050 \pm 0.0067 \pm 0.0130,
\label{delta0Final}
\eeq
where the first error is the estimate from the spread of the PAs and the
second error is due to the uncertainty  in $\alpha_s$. 

This result
can be compared with the  Borel model of Ref.~\cite{BJ08}  which gives (with  $\alpha_s(m_\tau^2) = 0.316 \pm 0.010$)
\beq
\delta_{\rm BM}^{(0)} = 0.2047 \pm 0.0029 \pm 0.0130 \qquad \mbox{(from Ref. \cite{BJ08})},
\label{delta0Borel}
\eeq
and with the estimate based on the optimal $C$-scheme
\beq
\delta_C^{(0)} = 0.2047 \pm 0.0034 \pm 0.0133  \qquad \mbox{(from Ref. \cite{BJM16})}.
\label{delta0Cscheme}
\eeq
In Eq.~(\ref{delta0Cscheme}) the first uncertainty is due to the truncation of the asymptotic series, while
in Eq.~(\ref{delta0Borel})  it arises from variations of the input value of the coefficient $c_{5,1}$ within their assumptions. In all cases
the second uncertainty stems from $\alpha_s$. The striking similarity of these three results is very remarkable, since they are obtained
from independent methods. In our case, the known series coefficients are the only input used to construct PAs and predict the behaviour
of the series at higher orders. In the case of Ref.\cite{BJ08}, the Adler function is modelled using the available knowledge about the leading
renormalon singularities, while the result from the $C$-scheme is based on a renormalization scheme choice which
uses optimally the available information in the spirit of an asymptotic series. The latter can also be considered as model independent since
no assumption about higher orders or the renormalon structure of the Adler function is made. 
More recently, two other analyses appeared in which the value of $\delta^{(0)}$ is obtained. In Ref.~\cite{PMC2018}, using the
the Principle of Maximum Conformality (PMC) the result
\beq
\delta^{(0)}_{\rm PMC} = 0.2035 \pm 0.0030 \pm 0.0123  \qquad \mbox{(from Ref. \cite{PMC2018})}\label{delta0PMC}
\eeq
is obtained, where the first error is again from higher orders and the second from $\alpha_s$. This result is also in very good agreement with ours. 
In Ref.~\cite{Caprini18}, through  the use  $C$-scheme variations together with an optimal conformal mapping  (CM) that relies on the location of the leading renormalons 
the value
\beq
\delta_{\rm CM}^{(0)} = 0.2018 \pm 0.0211 \pm 0.0123  \qquad \mbox{(from Ref. \cite{Caprini18})}
\label{delta0conformalC}
\eeq
is found, in which the first uncertainty is from the truncation of the asymptotic
series and the second from the strong coupling. This result is also nicely compatible with the others quoted in this section.

\begin{figure}[!t]
\begin{center}
\includegraphics[width=.95\columnwidth,angle=0]{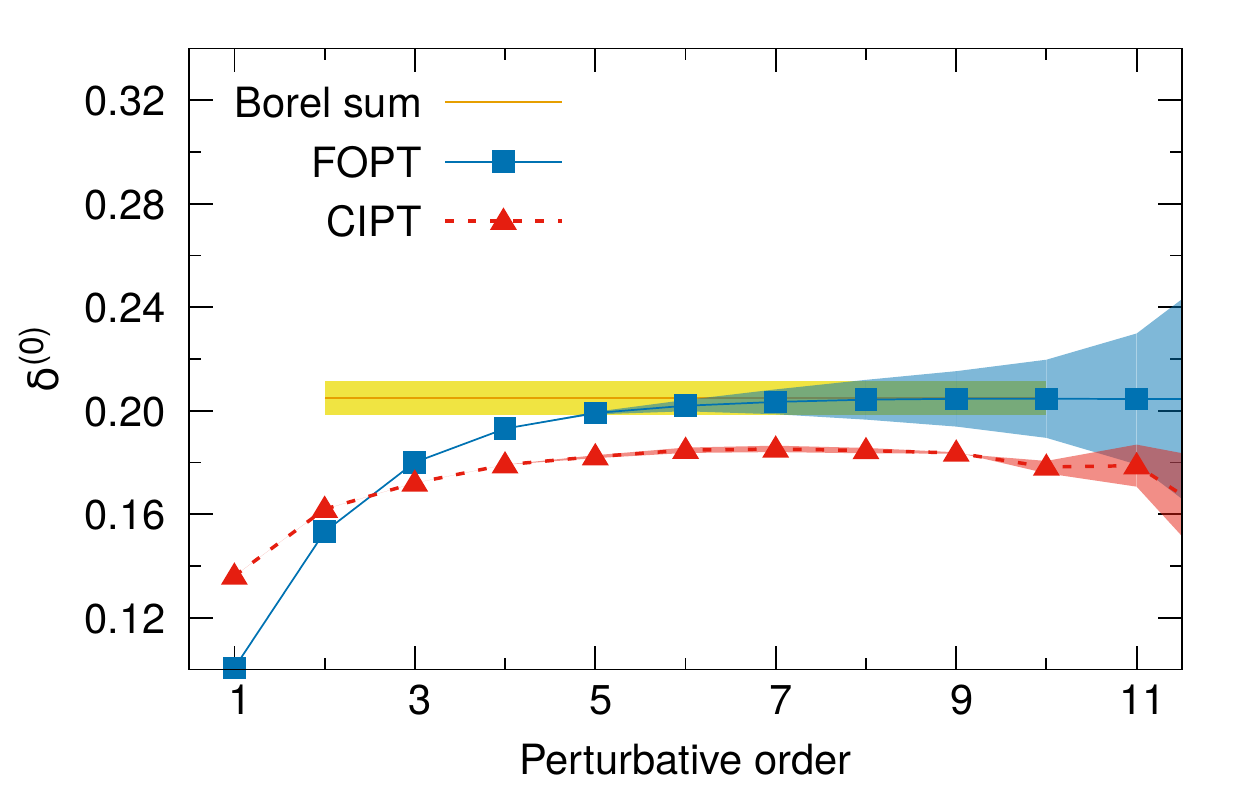}
\caption{Final results for  $\delta^{(0)}$ in QCD using the coefficients of Tab.~\ref{FinalCoeff} and the result of Eq.~(\ref{delta0Final}). The bands in the perturbative expansions reflect the uncertainty in the coefficients while the band in the sum of the series is obtained from the spread of the values from individual PAs (last columns of Tabs.~\ref{Delta0PadesQCD} and~\ref{Delta0BorelPadesQCD}). We use $\alpha_s(m_\tau^2) = 0.316.$ }
\label{delta0FinalPlot}
\end{center}
\end{figure}

With the coefficients of Tab.~\ref{FinalCoeff} we are finally in a
position to plot, in Fig.~\ref{delta0FinalPlot}, the perturbative
expansions of $\delta^{(0)}$ and compare them with the true value of the
series obtained from Eq.~(\ref{delta0Final}). The bands in the
perturbative expansions of Fig.~\ref{delta0FinalPlot} represent the
uncertainty from the series coefficients, given in
Tab.~\ref{FinalCoeff}, while the band in the Borel sum of the series
is the first error Eq.~(\ref{delta0Final}).
The uncertainties we are able to obtain from the optimal Pad\'e approximants
allow us to conclude that FOPT is the favored renormalization-scale setting
procedure  in the case of full QCD. The CIPT series, even though it looks
more stable around the fourth order, does not approach well the central value of the
sum of the series. The recommendation that FOPT is the best procedure in QCD
was advocated in Ref.~\cite{BJ08} in the renormalon-model context. Here it is
reobtained  in a  model-independent way.


\section{Conclusions}
\label{sec:conclusions}

In this work we have performed a systematic study of the Adler
function and of the perturbative QCD correction to the hadronic $\tau$
decay width, $\delta^{(0)}$, using Pad\'e approximants and its
variants. We have used the large-$\beta_0$ limit of QCD, where the
series are known to all orders in $\alpha_s$, as a laboratory to test
our strategy. We were able to show that the method always works
provided a large enough number of coefficients is known. Since in QCD
only the first four have been calculated exactly, we have devised
strategies with the aim of accelerating the convergence of the
approximants. The success of these strategies can be understood in
terms of the analytic structure of the Borel transformed series.  The
model independent acceleration methods simplify this structure either
by suppressing the residue of some poles or by reducing their
multiplicity.

A similar suppression of the poles is also found in the
Borel transform of $\delta^{(0)}$ which automatically leads to a more
regular series that is more amenable to approximation by PAs. We have
exploited this fact to show that, in large-$\beta_0$, the PAs formed
to the $\alpha_s$ expansion of $\delta^{(0)}$ and the D-log PAs constructed to its Borel
transform $B[\delta^{(0)}]$ are the optimal model-independent way of
extracting the higher-order coefficients of the Adler function.

We have then applied the same procedure to full QCD. The excellent
``postdiction'' of the coefficient $c_{4,1}$, which is known since
2008~\cite{BCK08}, as well as the striking stability of the results
gives us confidence that the method also works in QCD.
From PAs to the fixed-order expansion of  $\delta^{(0)}$ and D-log PAs
to $B[\delta^{(0)}]$ we extract the final results of this paper, given
in Tab.~\ref{FinalCoeff} and Eq.~(\ref{delta0Final}). These results
allow us to reconstruct very reliably the perturbative expansions of $\delta^{(0)}$
up to at least the tenth order. Finally, Fig.~\ref{delta0FinalPlot} shows that  our model-independent
reconstruction of the higher-order series coefficients favours FOPT as
the best procedure to set the renormalization scale at and around the $\tau$ mass.

We should remark that our final results are similar to the
model-dependent reconstruction of the series put forward in
Ref.~\cite{BJ08} and further discussed in Ref.~\cite{BBJ13}. This
lends  support to renormalon model used in these works, even though
the use of PAs show that one should be careful when interpreting the
parameters of such models. In the case of the renormalon models, the
analytic structure is completely fixed, the only freedom is left to
the residues. Therefore, we should expect that these are ``effective
residues'', in the spirit of Pad\'e-type approximants, and can only be
compared with the true residues in a hierarchical way, since they must account
for the infinite tower of poles that  must
be mimicked by the model.

Apart from providing reliable estimates for the higher-orders
coefficients and indicating that FOPT is preferred, our results could
be the basis for an $\alpha_s$ extraction based on the Borel sum of
$\delta^{(0)}$. The fact that the results of Eqs.~(\ref{delta0Final}--\ref{delta0conformalC}) are so close suggest
that it may be realistic to do so. We also intend to investigate
further the analytic structure of $B[\delta^{(0)}]$ in QCD, since the
non-trivial result of Eq.~(\ref{BorelDelta0}) plays a central role in
our analysis. The simplicity and flexibility of the
method here developed suggests it could also be used to further explore 
non-perturbative contributions in the context of $\alpha_s$
determinations.

\section*{Acknowledgements}

We thank I. Caprini for comments regarding Ref.~\cite{Caprini18}.
DB thanks the kind hospitality of IFAE and the Universitat Aut\`onoma
de Barcelona where part of this work was carried out. This work was
partially supported by the S\~ao Paulo Research Foundation (FAPESP)
grants 2015/20689-9 and 2016/01341-4 and by CNPq grant 305431/2015-3.
The work of P.M. is supported by the Beatriu de Pin\'os postdoctoral
programme of the Government of Catalonia's Secretariat for
Universities and Research of the Ministry of Economy and Knowledge of
Spain.

\end{document}